\documentclass[11pt]{article}
\usepackage[mathcal,mathscr]{eucal}
\usepackage{latexsym,amssymb}
\usepackage{amssymb}
\def\beas{\begin{eqnarray*}}   
\def\eeas{\end{eqnarray*}} 
\def\PHI{{\widehat {\Phi}}}
\def\K{\varkappa}

\newcommand{\be}{\begin{equation}} 
\newcommand{\ee}{\end{equation}} 
\newcommand{\ba}{\begin{eqnarray}} 
\newcommand{\ea}{\end{eqnarray}} 
\newcommand{\pa}{\partial} 
\renewcommand{\Re}{\mathrm {Re} \,}
\renewcommand{\Im}{\mathrm {Im} \,}

\def\bea{\begin{eqnarray}} 
\def\eea{\end{eqnarray}} 
\def\le{\left} 
\def\ri{\right} 
\def\l{\lambda}

\def\bes{$$} 
\def\ees{$$}

\title{\bf AdS/CFT correspondence for $n$-point functions}
\author{ Marco Bertola$^{\rm a}$,
         Jacques Bros$^{\rm c}$, 
         Ugo Moschella$^{\rm b,c}$, 
         Richard Schaeffer$^{\rm c}$\\[3pt]
      $^{\rm a}$ {\small {CRM, Universite' de Montreal, Montreal (Quebec), H3C 3J7}}  \\
       $^{\rm b}$ {\small  Dipartimento di Scienze Matematiche Fisiche e  
                   Chimiche,}\\
       {\small Universit\`a dell'Insubria, 22100 Como and INFN sez. di Milano, Italy} \\
       $^{\rm c}$ {\small Service de Physique Th\'eorique, C.E. Saclay,
91191 Gif-sur-Yvette, France} } 

\begin{document}
\maketitle

\abstract

We provide a new general setting for scalar interacting fields on the
covering of a $d+1$--dimensional 
AdS spacetime.  
The formalism is used at first to construct a one-parameter 
family of field theories, each living on a corresponding 
spacetime submanifold of AdS,   
which is a cylinder ${\mathbb R}\times {{\mathbb S}_{d-1}}$.  
We then introduce a  limiting procedure which
directly produces L\"uscher-Mack CFT's on the covering of the AdS
asymptotic cone.  Our AdS/CFT correspondence  
is generally valid for interacting
fields, and is illustrated by a complete treatment of two-point functions,
the case of Klein-Gordon fields appearing as particularly simple in our
context.

We also show how the Minkowskian representation of these boundary CFT's
can be  directly generated by an alternative limiting procedure 
involving Minkowskian theories in horocyclic sections 
(nowadays called ($d-$1)--branes, 3--branes for $AdS_5$).
These theories are restrictions to the brane 
of the ambient AdS field theory considered.
This provides a  more general correspondence between the AdS field theory
and a Poincar\'e invariant QFT on the brane, satisfying 
all the Wightman axioms.
The case of two-point functions is again studied in detail from this viewpoint
as well as the CFT limit on the boundary.


\newpage

\section{Introduction}

Shortly after the appearance of Maldacena's AdS/CFT conjecture
\cite{Maldacena:1997re}, a proposal to make it effective has been
formulated independently in \cite{Gubser:1998bc} and
\cite{Witten:1998qj}, and then discussed   in a  
large number of papers 
(for a review see \cite{Aharony:1999ti}).
The proposal prescribes a way  to compute the Green's functions of the
boundary CFT in terms of the Euclidean classical supergravity action of
an AdS field configuration which obeys to precise boundary conditions.
Most of these works are thus making use of classical field theory methods 
(mainly in a Euclidean formulation allowing the use of Dirichlet's 
problem but also   
in a Lorentzian formulation in terms of modes of the Klein-Gordon equation   
\cite{Balasubramanian:1998sn}).  

However, since the ideas of the AdS/CFT conjecture suggest the
existence of a rich and still uninvestigated conceptual content at the
level of quantum fields, it is natural that more recent works in this
domain have made use of the already existing (and more than two decades
old) results about quantum field theory (QFT) on the AdS spacetime.  In
this connection, one must quote the pioneering approach of \cite{Avis},
whose main concern was to specify boundary conditions such that the
difficulties arising  by the lack of global hyperbolicity of the
underlying AdS manifold could be circumvented and the resulting QFT be
well defined.  Another, earlier, approach was also given on the basis
of group-theoretical methods \cite{Fronsdal:1974} following ideas that
can be traced back to Dirac \cite{Dirac}.  

Both of these approaches
have influenced very much the recent research on the AdS/CFT subject.
However, their applicability is more or less limited to free AdS QFT's
(even if they can produce useful ingredients for  perturbative
calculations) and there is the need of setting the AdS/CFT debate on a
more general basis\footnote{In this spirit such a setting has been
proposed \cite{rehren} in the general framework of algebras of
local observables (or ``local quantum physics'' in the sense of R. Haag
\cite{haag}).}  in which both AdS quantum fields and boundary CFT's
would be treated from the viewpoint of the structural properties of
their $n-$point correlation functions.  

In a basic work by L\"uscher
and Mack \cite{Luscher:1975ez}, the concept of global conformal
invariance in Minkowskian QFT has been associated in a deep and
fruitful way with the general framework of QFT on the covering of a
quadratic cone with signature $(+,+,-,\cdots,-)$ (in one dimension
more). Since such a cone is precisely the asymptotic cone of the AdS
quadric, it seems quite appropriate to try to formulate the AdS/CFT
correspondence in a way which exhibits as clearly as possible the
connection between the previous conceptual QFT framework on the cone
with a similar QFT framework on the AdS quadric. This is the first
purpose of the present work.

In Section 2 we introduce such a general framework for the study of
quantum fields on a ($d$+1)--dimensional AdS spacetime $AdS_{d+1}$ (or
more precisely on its covering) in the spirit of \cite{Streater}.
Then,  in Section 3, we present a direct and simple method for
obtaining correlation functions of {\em corresponding} conformal fields
on the boundary, method which does not rely on the use of any field
equation.  General interacting QFT's on the (covering of the) AdS
spacetime are assumed to satisfy a set of basic properties such as
locality, AdS covariance and a certain energy spectrum condition
formulated in terms of appropriate analyticity properties of the
$n$-point functions.  Together with these general properties, it is
also crucial to require a certain type of asymptotic behavior for the
$n$-point functions in terms of an {\em asymptotic dimensionality
condition}. The possibility to use such type of asymptotic condition
has  been mentioned  in a perturbative context in \cite{Banks:1998dd}.
We then introduce a limiting procedure which directly produces CFT's
(in the sense of \cite{Luscher:1975ez}) on the  covering of the
asymptotic cone of AdS spacetime; the conformal covariance of the
corresponding Minkowskian (interacting) field theory on the boundary
is then proved without making use of any field equation.

Our approach therefore differs from the original largely followed
proposals of \cite{Gubser:1998bc} and \cite{Witten:1998qj} in the sense
that, rather solving explicitly for a field that is defined by
suitable boundary conditions on the edge -- which can be done in a
tractable way only for free fields -- we define a limiting procedure
that can be applied directly to the AdS correlation functions.  
This is much simpler already for free fields, but our treatment also
shows how to deal with {\it interacting fields} $AdS_{d+1}$ and we gain
a general non-perturbative foundation for the AdS/CFT conjecture,
formulated in terms of the $n-$point correlation functions of such
fields and treated from a model-independent point of view.

In Section 4 we provide a complete treatment of two-point correlation
functions.  By applying the  general setting of Section 2 we are able
to exhibit strong analyticity properties of  AdS two-point functions
\cite{Bros:1999}, which are closely similar to those enjoyed by
two-point functions in flat spacetime or in de Sitter spacetime
\cite{Bros:1994dn,Bros:1996js}.  In the case of Klein-Gordon fields,
these analyticity properties fix completely their form 
to be  necessarily  appropriate second-kind Legendre
functions, as obtained (for the four-dimensional case) in the
group-theoretical approach of \cite{Fronsdal:1974}; the CFT limits of
such two-point functions are then directly computable in full
consistency with the given general formulation of the AdS/CFT
correspondence.

We also provide an alternative construction based
on the Poincar\'e or horocyclic coordinate system for $AdS_{d+1}$.
This way of looking to the AdS spacetime as a
warped manifold with Poincar\'e sections (today called $d-$1-- branes) 
has recently gained an enormous interest in a phenomenological 
and cosmological context \cite{Randall:1999ee}.

We show that by restriction, 
AdS correlation functions satisfying our general properties of Section 2 
define acceptable two-point  Minkowskian QFT
correlation functions on the branes 
(corresponding to
flat $d$-dimensional 
spacetime sections of the AdS manifold).  
In particular, in Section 5 we establish a spectral decomposition  for
the  AdS Klein-Gordon fields  naturally associated to the
Poincar\'e  foliation of the AdS quadric.  In this
scheme the well-known ambiguity for low mass theories
\cite{Breitenlohner:1982jf} is seen to be linked to the lack of
essential self-adjointness of the Bessel's differential operator
\cite{Titchmarsh:1962} which naturally arises in the Poincar\'e
coordinates.

In the last part of our paper, we introduce a more general possible use
of the Poincar\'e foliation by showing how general Minkowskian
interacting QFT's can be produced by taking the restrictions of general
AdS QFT's to the branes.  Moreover, by sending these leaves to infinity
we exhibit a limiting procedure for the Minkowskian QFT's in the leaves
which provides an alternative presentation of the previously defined
AdS/CFT correspondence. The interest of this presentation is that it is
entirely expressed in terms of Minkowskian theories satisfying the
Wightman axioms, and can be of interest for the multidimensional
approach to phenomenology and cosmology \cite{Randall:1999ee}.

\section{General QFT in AdS spacetime}
\subsection{Notations and geometry}
We consider  the vector space ${\mathbb R}^{d+2}$ equipped with the
following pseudo-scalar product:
\begin{equation}
X\cdot X'  = {X^0} {X'}^0 - {X^1} {X'}^1 - \cdots - {X^d}{X'}^d   
+ X^{d+1}  {X'}^{d+1}\ .
\label{ambientmetric}
\end{equation}
The $(d+1)$-dimensional AdS universe can then be identified with the 
quadric  
\begin{equation}
AdS_{d+1} = \{ X \in \mathbb R^{d+2},\;\; {X^2}=R^2\}, 
\end{equation}
where  $X^2= {  X} \cdot{X}$, endowed with the induced metric
\begin{equation}
{\mathrm d}s^2_{AdS} = \left.\left(d{{X}^0}^{\,2}-d{{X}^1}^{\,2} 
- \cdots + d{{X}^{d+1}}^{\,2}\right) \right|_{AdS_{d+1}}.
\label{metric}\end{equation}
The AdS relativity group  is   $G =SO_0(2,d )$, that is   
the component connected to the identity of 
the pseudo-orthogonal group $O(2,d )$.
Two events  $X$, $ X'$ of $AdS_{d+1}$ 
are space-like separated if $(X- X')^2<0$, 
i.e. if  $X\cdot  X'>R^2$.

We will also consider the complexification of $AdS_{d+1}$:
\begin{equation}
AdS^{(c)}_{d+1} = \{ Z = X+iY \in \mathbb C^{d+2}, \;\;Z^{2}= R^2\}. 
\end{equation}
In other terms, $Z = X+iY$ belongs to $AdS^{(c)}_{d+1}$ if and only if
$X^2 - Y^2 = R^2$ and $X\cdot Y = 0$. In the following we will put for notational simplicity $R=1$.\\[5pt]
 
We shall make use of two parametrizations for  the AdS manifold.\\[5pt]
The {\sl   ``covering parametrization'' 
$X = X[r,\tau, {\rm e}]$:} it is obtained by intersecting $AdS_{d+1}$ 
with the cylinders with equation  $ \{{X^{0}}^2 + {X^{d+1}}^2  = r^2 + 1\}$, and is given by
\begin{equation}
\left\{\begin{tabular}{lclcll}
$X^{0} $  &=& $\sqrt{r^2 + 1} \sin \tau $ & \cr
$X^{i} $  &=& $r  {\rm e}^i $                   & ${ i=1,...,d}$ & \cr
$X^{d+1}$ &=&$ \sqrt{r^2 + 1} \cos \tau $& 
\label{sphericcoordinates}
\end{tabular}\right.
 \end{equation}
with ${\rm e}^2 \equiv {{\rm e}^1}^2 + \ldots + {{\rm e}^d}^2 = 1$ and $r \ge 0$. 
For each fixed value of $r$, the corresponding ``slice'' 
\begin{equation}
C_{r}=  {AdS}_{d+1} \cap  \{{X^0}^2+
{X^{d+1}
}^2  = r^2 + 1\}
\label{CR}
\end{equation}
of 
$AdS_{d+1}$ is a manifold ${\mathbb S}_1 \times {\mathbb S}_{d-1}$.  
The complexified space $AdS^{(c)}_{d+1}$ is obtained by giving arbitrary complex values to 
$r, \tau $ and to the coordinates ${\rm e}
 = ({\rm e}^i)$ on the unit $(d-1)-$sphere.

The parametrization (\ref{sphericcoordinates}) allows one to introduce relevant coverings  
of $AdS_{d+1}$ and ${AdS}^{(c)}_{d+1}$ by unfolding 
the $2 \pi -$periodic coordinate $\tau$ (resp. $\Re{\tau}$), interpreted as a time-parameter:
these coverings are denoted respectively by
$\widehat {AdS}_{d+1}$ and $\widehat {AdS}^{(c)}_{d+1}$. 
A privileged ``fundamental sheet'' is defined on these coverings by 
imposing the condition $-\pi < \tau < \pi$
(resp. $-\pi < \Re{\tau} < \pi$). 
This procedure also associates with each manifold $C_r$ its covering $\hat
C_r$ 
which is a cylinder ${\mathbb R}_{\tau} \times {{\mathbb S}_{d-1}}_{\rm e}$.  
We will use the symbols $X$, $Z$, ..., also to denote points of the coverings.

Similarly one introduces a covering $\hat G$ of the group $G$ by taking in 
$G$ the universal covering of the rotation subgroup in the $(0,d+1)-$plane.
By transitivity, 
$AdS_{d+1}$ and $\widehat {AdS}_{d+1}$ are respectively generated by the action
of $G$ and $\hat G$ on the  base point  
$B=(0,\ldots,0,1)$ . 

The physical reason which motivates the introduction of the covering  
$\widehat {AdS}_{d+1}$, that is the requirement of nonexistence of closed
time-loops, also leads us  
to specify the notion of space-like separation in 
$\widehat {AdS}_{d+1}$ as follows: 
let  $X,X'\in \widehat{AdS}_{d+1}$ 
and let  $g$ an element of $\hat G$ such that
$X' = gB$; define  $X_g= g^{-1} X$.\\
$X$ and $X'$ are spacelike separated if $X_g$ is in the fundamental sheet of
 $\widehat {AdS}_{d+1}$ and  $(X-X')^2 \equiv (g^{-1}X-g^{-1}X')^2 <0$.
 This implies that $X_g =X_g[r, \tau,{\rm e}]$ with $-\pi < \tau < \pi$ and 
$ \sqrt{r^2 + 1} \cos \tau >1 $.\\
It is also interesting to note that on each manifold $C_r$ the condition of
space-like
separation between two points $X =X[r,\tau,{\rm e}]$ and $X' = X'[r,\tau',
{\rm e}']$  
reads (in view of (\ref{sphericcoordinates}):
\begin{equation}
(X-X')^2 = 2(r^2 +1)(1 -\cos (\tau -\tau')) - r^2 ({\rm e}-{\rm e}')^2 <0,
\label{causal}
\end{equation}
and that the corresponding covering manifold $\hat C_r$ therefore admits a global causal ordering
which is specified as follows:
\begin{equation}
(\tau,{\rm e}) > (\tau',{\rm e} ')\ \ \ {\rm  iff}\  \ \ \tau-\tau' > 
2 {\mathrm {Arcsin}} \left(\frac{({\rm e} -{\rm e}')^2} {4}  
\frac {r^2}{r^2+1}\right)^{\frac{1}{2}}. 
\label{timeorder}
\end{equation}

\vskip 0.4cm
\noindent
The {\sl  ``horocyclic parametrization'' $X = X(v,x)$:} 
it only covers the part $\Pi$  
of the AdS 
manifold which belongs to the half-space $\{X^d + X^{d+1}>0 \}$ of the ambient space 
and is obtained by intersecting $AdS_{d+1}$ with 
the hyperplanes $ \{X^{d} + X^{d+1} = e^ {v} =\frac{1}{u}\}$\footnote{The
 coordinate  ${u}=e^{- {v}}$ is frequently
called $z$ in the recent literature. 
We are forced to change  this notation because we reserve the letter 
$z$ to complex quantities. By allowing  also negative values for $u$
the coordinate system (\ref{coordinates}) 
covers almost all the real manifold $AdS_{d+1}$.},  
each slice $\Pi_v$ (or ``horosphere'') being an hyperbolic paraboloid: 
\begin{equation}
\left\{\begin{tabular}{lclcll}
 $X^{\mu} $&=& $e^{ {v}} x^\mu  $ & =& $\frac{ 1}{u}x^\mu  $& $
 { \mu=0,1,...,d-1}$\cr 
 $X^{d} $&=& $\sinh  {v} + \frac 12 e^{ {v}} x^2 $  & =& $\frac{1-u^2}{2u} + 
\frac {1}{2 u} x^2$ & $x^2 ={ x^0}^2- {x^1}^2- \cdots -{x^{d-1}}^2$  \cr
 $X^{d+1}$&=&$ \cosh  {v} - \frac 12 e^{ {v}} x^2$ &  =& $\frac{1+u^2}{2u} - 
\frac {1}{2 u} x^2$& 
\label{coordinates}
\end{tabular}\right.
 \end{equation}
In each slice $\Pi_v$, $x^0,...,x^{d-1}$ can be seen as coordinates of an event of a $d$-dimensional
Minkowski spacetime ${\mathbb M}^{d}$ with metric $ds^2_{{M}}= d{{x^0}}^{\,2}-
d{{x^1}}^{\,2} - \ldots -d{{x^{d-1}}}^{\,2}$ 
(here and in the following where it appears, an index {\em {\small M}} stands for Minkowski).  
This explains why  the  horocyclic coordinates $(v,x)$ of the 
parametrization   (\ref{coordinates}) 
are also called  Poincar\'e coordinates.
The scalar product (1) and the AdS metric can then be rewritten as follows: 
\begin{eqnarray}
&& X\cdot X' = \cosh( {v}- {v} ')  - \frac 12 e^{ {v}+ {v}'} \le(x-x'\ri)^2,
\label{7}\\
&& {\mathrm d}s^2_{AdS} = e^{2 {v}} {\mathrm d}s^2_{{M}}-{\mathrm d} {v}^2 
= \frac{1}{u^2}({{\mathrm d}s^2_{{M}}-{\mathrm d} u^2}). \label{metric1}
\end{eqnarray}
 Eq. (\ref{7}) implies that 
\begin{equation}
(X(v,x)-X(v,x'))^2 = e^{2v}(x-x')^2.
\label{lll}
\end{equation}
This in turn implies that space-like separation in any slice $\Pi_v$ can be understood
equivalently in the Minkowskian sense of the slice itself 
or in the sense of the ambient 
AdS universe.

Eq. (\ref{metric1}) exhibits the region $\Pi$ of $AdS_{d+1}$  
as a warped product \cite{O'Neill} with warping function 
$\omega( {v})=e^{ {v}}$ and fibers conformal to 
${\mathbb M}^{d}$. 
The use of this parametrization is crucial in a recent 
approach to the mass hierarchy problem  
\cite{Randall:1999ee} and to multidimensional cosmology.
In this context the slices $\Pi_v$ are called branes.
Finally, the representation of $\Pi$ by the parametrization (\ref{sphericcoordinates}) is specified by
considering $\Pi$ as embedded in the fundamental sheet of 
$\widehat {AdS}_{d+1}$; it is therefore described by the following conditions
on the 
coordinates $r, \tau, {\rm e}$:
\begin{equation}
-\pi < \tau< \pi; \ \ \  r{\rm e}^d +  
\sqrt{r^2 + 1} \cos \tau >0 
\end{equation}

\vskip 0.4cm
\noindent
The {\sl  ``Euclidean'' submanifold ${E_{d+1}}$ of 
$\widehat {AdS}_{d+1}^{(c)}$}   
 is the set of all points $Z=X+ i Y$ in 
$\widehat {AdS}_{d+1}^{(c)}$ such that $X= (0,X^1,...,X^{d+1})$,
$Y= (Y^0,0,...,0)$ and $X^{d+1} >0$. It is therefore represented by the 
upper sheet (characterized by the condition $X^{d+1} >0$) of the two-sheeted hyperboloid with equation
${X^{d+1}}^2 - {Y^0}^2 -{X^1}^2- \cdots -{X^d}^2 =1$. 
${E_{d+1}}$ is equally well represented in both parametrizations (\ref{sphericcoordinates}) and (\ref{coordinates}) as follows: 
\begin{equation}
Z = Z[r,\tau = i\sigma, {\rm e}];\ \ (r,\sigma,{\rm e} ) \in 
{\mathbb R} \times  
{\mathbb R} \times 
{{\mathbb S}_{d-1}}
\label{10}
\end{equation}
or
\begin{equation}
Z= Z(v, (iy^0, x^1,...,x^{d-1}));\ \ v\in
{\mathbb R},\   
(y^0, x^1,...,x^{d-1}))\in  
{\mathbb R}^{d}. 
\label{11}
\end{equation}
In view of (\ref {10}),  
${E_{d+1}}$ is contained in the fundamental sheet of 
$\widehat {AdS}_{d+1}^{(c)}$. 

For each $v$, the complexification $\Pi_v^{(c)}$ of the horosphere $\Pi_v$
is parametrized by formulae (6) in which $x$ is replaced by the complex
Minkowskian vector $z = x + iy = (z^0,...,z^{d})$; the Euclidean
submanifold of this complex Minkowskian manifold
is obtained as the  intersection   $\Pi_v^{(c)}\cap {E_{d+1}}$.

\vskip 0.4cm
\noindent

\subsection{Quantum Field Theory}
Let us consider now a general QFT on $\widehat {AdS}_{d+1}$; for simplicity we
limit the present discussion to one scalar field $\Phi(X)$. 
According to the general reconstruction procedure   \cite{Streater}, a theory 
is completely determined by the set of all  
$n$-point vacuum expectation values (or ``Wightman functions'')  of  the field $\Phi$,
given as distributions on the corresponding product manifolds
$(\widehat {AdS}_{d+1})^n$: 
\begin{equation}
{\cal W}_{n}(X_1, \ldots X_n)
=\langle \Omega, \Phi(X_1) \ldots \Phi(X_n) \Omega \rangle.
\label{npoint}
\end{equation}
These distributions are supposed to be tempered
when represented in the variables of the covering parametrization 
$X_j = X_j[r_j,\tau_j,{\rm e}_j]$  and to satisfy a set of 
general requirements 
which we will  specify below. 

Before doing it, we remark that a QFT on
$\widehat {AdS}_{d+1}$ can be projected to a theory on  
$ AdS_{d+1}$ itself if it is $2 \pi-$periodic in the time-parameter $\tau$,
namely if each 
${\cal W}_{n}(X_1, \ldots X_n)$ is invariant under all individual
substitutions  
$X_j[r_j,\tau_j,{\rm e}_j] \to  
X_j[r_j,\tau_j + 2\pi,{\rm e}_j]$.   

An important class of fields, which can be explicitly constructed in a 
Fock space, is 
the class of ``generalized free fields''; these fields are  
completely determined by their two-point
function ${\cal W}_{2}(X_1, X_2)$. In particular, the 
Klein-Gordon fields are those for which ${\cal W}_{2}(X_1,X_2)$ satisfies the 
corresponding field equation w.r.t. both
points.  Of course there are in general infinitely many inequivalent
solutions to this problem (encoded in the choice of ${\cal W}_2$) 
and one has to select 
the meaningful ones on the basis of some physical principle;
the existence of many possible theories
even for a free field of a given mass is no surprise.  

We shall assume that the distributions ${\mathcal W}_n$ satisfy the following
properties: 
{\em AdS invariance, positive-definiteness, hermiticity,
local commutativity, analyticity corresponding to an appropriate spectral condition
and ``dimensional boundary conditions'' at infinity}.\\[10pt]
The requirement of {AdS invariance} (corresponding to the 
scalar character of the field)   can be written as follows:
\begin{equation}
{\cal W}_{n}(gX_1, \ldots gX_n) =  {\cal W}_{n}(X_1, \ldots X_n)
 \qquad \mbox{for any } g\in {\hat G}.
\label{covariance}
\end{equation}
The usual positivity 
and hermiticity properties \cite{Streater} are 
valid for scalar QFT's on any spacetime and we do not spell them out.

\vskip 0.4cm
\noindent
a) {\em Local commutativity.}
$\Phi(X)$ commutes (as an operator-valued distribution) with $\Phi(X')$ for $X,X'$ space-like separated 
in the sense of the covering space 
$\widehat {AdS}_{d+1}$, 
as defined above  
(for theories in  
$ AdS_{d+1}$ itself, it implies commutativity under the only condition 
$(X-X')^2 <0$, which then includes a certain condition of commutativity at  
periodic time-like separations). 
As in the Minkowskian case, this postulate is equivalent to the coincidence of 
permuted Wightman functions at space-like separation of consecutive arguments
$X_j, X_{j+1}$ \cite{Streater}.

\vskip 0.4cm
\noindent
b) {\em Analyticity corresponding to energy spectrum  condition.}
Since the parameter of the covering group of the rotations in the $(0,d+1)-$plane 
is interpreted as a genuine time-translation for the observers in all the corresponding
Killing trajectories, and since the complexifications of these trajectories do not 
exhibit any geometrical periodicity\footnote{Such geometrical periodicity in purely imaginary times gives rise to thermal effects 
for the corresponding observers, 
as it has already been checked in various examples of QFT on curved spacetimes \cite{Hawking:1974df,Gibbons:1977mu,Sewell,Bros:1996js}.} 
in $\widehat {AdS}_{d+1}^{(c)}$, 
it is legitimate to consider QFT's for which the corresponding infinitesimal generator $J_{0,d+1}$ 
is represented by a self-adjoint  operator whose spectrum is 
bounded from below: the latter postulate is in fact interpretable 
as a reasonable spectral condition on the energy, 
valid for all these observers.
By using the standard Laplace transform argument \cite{Streater,Reed} 
in the 
corresponding time-variables $\tau_1,...,\tau_n,$  
one is led to formulate this spectral condition by the following analyticity property
of the Wightman functions:
\\[10pt]
{\em Each tempered distribution ${\cal W}_n(X_1[r_1,\tau_1,{\rm
e}_1],...,X_n[r_n,\tau_n, {\rm e}_n])$ 
is the boundary value of a
holomorphic function ${W}_n (Z_1,...,Z_n)$ which is defined in a complex neighborhood 
of the set $\{Z=(Z_1,...,Z_n); Z_j=X_j+iY_j \in {\widehat {AdS}_{d+1}^{(c)}};$
$ Z_j = Z_j[r_j,\tau_j, {\rm e}_j];
\ $ $ \Im {\tau_1} <\Im{\tau_2} <\cdots <\Im{\tau_n}\}$.}\\[10pt] 
As a by-product, the Schwinger function $S_n$, that is 
 the restriction of each ${W}_n$ to the Euclidean submanifold
$\{(Z_1,...,Z_n) \in {({E_{d+1}})}^n;\  \sigma_1 <\sigma_2 <\cdots
<\sigma_n\}$, is well-defined.

\vskip 0.4cm
\noindent
c) {\em Dimensional boundary conditions at infinity.} 
In order to obtain relevant QFT's on the boundary of AdS spacetime (see
section 3), we are led to postulate a certain type of power-decrease at
infinity for the Wightman functions which we call ``dimensional
boundary conditions at infinity''; such conditions can be shown to be
valid in the case of Klein-Gordon fields (see section 4).   

By making use of the coordinates (\ref{sphericcoordinates}) 
we say that a QFT on 
$\widehat {AdS}_{d+1}$ is of asymptotic dimension $\Delta$ if the 
following limits exist 
in the sense of distributions:
 \begin{eqnarray}
 & \lim_{min(r_1,...,r_n) \to {+\infty}} &(r_1\cdots r_n)^{\Delta} 
{\cal W}_n (X_1[r_1,\tau_1,{\rm e}_1],...,X_n[r_n,\tau_n,{\rm e}_n]) \cr 
&&= {\cal W}_n^{\infty}([\tau_1,{\rm e}_1],...,[\tau_n,{\rm e}_n])
\label{15}
 \end{eqnarray}
\vskip 0.4cm
We have to show that the above condition is meaningful, since it is not
true in general that a distribution ${\cal W}_{n}(X_1, \ldots X_n)$ can be
restricted to the  submanifold $\prod_{j=1}^{n} \hat C_{r_j}$ of $(\widehat
{AdS}_{d+1})^n$ ($C_r$ was defined in Eq. (\ref{CR})).  Our spectral condition b) implies
that this can be done in the present framework.  In fact, for each
fixed $r_1,...,r_n$ and ${\rm e}_1,...,{\rm e}_n$, the existence of an analytic
continuation $W_n$ of ${\cal W}_n$ in the variables
$\tau_1,...,\tau_n$  of the covering parametrization
(\ref{sphericcoordinates}) in the tube domain $T_n =
\{(\tau_1,...,\tau_n);\ \Im {\tau_1} <\Im{\tau_2} <\cdots
<\Im{\tau_n}\}$ implies that the boundary value of ${ W}_n$ on the
reals from this tube is a distribution
in the variables $\tau_1,...,\tau_n$ {\it on each leaf obtained by fixing all the parameters 
$r_j$ and ${\rm e}_j$} and that it is even a regular (namely $C^{\infty}$)
function 
of all these leaf  
parameters. The limit in Eq. (\ref{15}) is therefore also defined as a distribution in the variables 
$\tau_1,...,\tau_n$ with $C^{\infty}$ dependence with respect to the variables
${\rm e}_j$. 
Moreover, it is then natural to assume that
{\em  the limit in Eq.(\ref{15}) can be extrapolated to
the holomorphic functions ${W}_n$ in their tube domains $T_n$}
so that the corresponding limits ${W}_n^{\infty}$ 
are themselves holomorphic in $T_n$ and admit the corresponding distributions 
$ {\cal W}_n^{\infty}$ as their boundary values on the reals.  
By restricting all these holomorphic functions to the Euclidean manifolds $\tau_j = i\sigma_j,\ j=1,...,n,$ 
one then obtains a similar condition for the Schwinger functions $S_n$ and the corresponding 
limits $S_n^{\infty}$.\\[10pt]  
If one wishes  to select QFT's satisfying the property of {\em uniqueness of the vacuum}, one should supplement 
the previous requirements by an appropriate cluster property on the $n-$point functions.
In order to obtain a relevant cluster property for the corresponding  
L\"uscher-Mack 
CFT on the cone $\widehat{\cal C}_{2,d}$ obtained by the procedure described in
our next section (namely the ``conformal cluster property'' described in \cite{Luscher:1975ez}),
one should formulate here a similar cluster property on the Schwinger
functions on  
$\widehat {AdS}^{(c)}_{d+1}$, namely:  
\begin{eqnarray}
&&\lim_{\sigma \to +\infty}  \;\;\;{\cal W}_{m+n} (X_1[r_1,i\sigma_1,{\rm
e}_1],\ldots,X_m[r_m,i\sigma_m ,{\rm e}_m], \cr
&& \;\;\;\;\;\;\;\;\;\;X_{m+1}[r_{m+1},i(\sigma_{m+1}+\sigma),{\rm
e}_{m+1}],\ldots,
X_{m+n}[r_{m+n},i(\sigma_{m+n}+\sigma),{\rm e}_{m+n}]) =  \cr 
&& \cr
&&  = {\cal W}_m (X_1[r_1,i\sigma_1,{\rm e}_1],...,X_m[r_m,i\sigma_m,{\rm
e}_m])  \times \cr 
&& \;\;\;\;\;\;\;\;\;\times {\cal W}_n (X_{m+1}[r_{m+1},i\sigma_{m+1},{\rm
e}_{m+1}],...,X_{m+n}[r_{m+n},i\sigma_{m+n},{\rm e}_{m+n}]).  
\label{cluster}
\end{eqnarray}\\[15pt]
{\bf Local Quantum Field Theories on the manifolds $ {\hat  C_r} $}.\\[10pt] 
As a special application of the previous framework, it is meaningful to consider the restrictions
of the distributions ${\cal W}_n$ 
to the submanifolds $\left({\hat C_r}\right)^n$  of $\left({\widehat {AdS}_{d+1}}\right)^n$   
(i.e. to the case when all variables $r_j$ are equal to $r$). 
One then notices 
that the positivity conditions satisfied by assumption by the distributions 
${\cal W}_n$ on 
$\widehat {AdS}_{d+1}$ can be extended to test-functions 
of the variables $\tau_j$ and ${\rm e}_j$
localized in these submanifolds
$r_1 =\cdots = r_n =r$. In view of the standard reconstruction procedure
\cite{Streater}, this allows one to say that in each slice $\hat C_r$ the
given field on   
${\widehat {AdS}_{d+1}}$ yields by restriction a well-defined quantum field
$\Phi_r(\tau,{\rm e})$. 
This field is obviously invariant under the product of the translation group with time-parameter $\tau$ 
by the orthogonal group $SO(d)$ of space transformations  
acting on the sphere ${\mathbb S}_{d-1}$ of the variables ${\rm e}$. 
Moreover, it follows from the locality postulate a) together with
Eqs. (\ref{causal})  
and (\ref{timeorder}) that the field $\Phi_r$ also satisfies local
commutativity in the sense
of the spacetime manifold $\hat C_r$.
Finally, in view of b), the $n-$point functions of $\Phi_r$ are (for each $r$)
boundary values of 
holomorphic functions of the complex variables $\tau_1,...,\tau_n$ in the tube
$T_n$,  
which shows that these theories satisfy a spectral condition with respect to the generator
of time-translations. 

\section{Correspondence with  conformal field theories on $\widehat{\cal
C}_{2,d}$ {\em \`a la } L\"uscher-Mack} 

We shall  now introduce the asymptotic cone ${\cal C}_{2,d}$ (resp. 
${\cal C}^{(c)}_{2,d}$) of 
$AdS_{d+1}$ (resp. ${AdS}^{(c)}_{d+1}$) and wish to identify the limit (in the sense of
Eq. (\ref{15}))
of a QFT on  
$\widehat {AdS}_{d+1}$ satisfying the previous properties  
with a QFT on the corresponding covering   
$\widehat {\cal C}_{2,d}$ of 
${\cal C}_{2,d}$.   
To do this, we first notice that by adapting the covering parametrization (\ref{sphericcoordinates})
of $\widehat {AdS}_{d+1}$ to the case of its asymptotic cone, 
${\cal C}_{2,d}= \{ \eta =(\eta^0,...,\eta^{(d+1)});\ {\eta^0}^2- {\eta^1}^2- \cdots- {\eta^d}^2
+{\eta^{d+1}}^2 =0 \}, $ 
one readily obtains the
following parametrization  
(with the same notations as \cite{Luscher:1975ez}, but in
dimension $d+2$): 
\begin{equation}
\left\{\begin{tabular}{lclcll}
$\eta^{0} $  &=& $r \sin \tau $ & \cr
$\eta^{i} $  &=& $r  {\rm e}^i $                   & ${ i=1,...,d}$ & \cr
$\eta^{d+1}$ &=&$ r \cos \tau $& 
\label{conecoordinates} 
\end{tabular}\right.
\end{equation}
with ${{\rm e}^1}^2 + \ldots + {{\rm e}^d}^2 = 1$ and $r \ge 0$,   
or in brief: $\eta = \eta[r, \tau,{\rm e}]$.

The parametrization (\ref{conecoordinates}) allows one to introduce the
coverings  
$\widehat {\cal C}_{2,d}$ and $\widehat {\cal C}^{(c)}_{2,d}$ of  
${\cal C}_{2,d}$ and $ {\cal C}^{(c)}_{2,d}$.  
by again unfolding 
the $2 \pi -$periodic coordinate $\tau$ (resp. $\Re{\tau}$).  
A privileged ``fundamental sheet'' is defined on these coverings by 
imposing the condition $-\pi < \tau < \pi$
(resp. $-\pi < \Re{\tau} < \pi$). 

We also note that the standard condition of space-like separation on 
${\cal C}_{2,d}$
is similar to the condition chosen on the AdS spacetime, namely 
\begin{equation}
(\eta -\eta')^2 =
r^2\left[4\left(\sin\left(\frac{\tau-\tau'}{2}\right)\right)^2 - ({\rm e}-{\rm
e}')^2\right]
= -2 r^2( \cos(\tau - \tau') - {\rm e}\cdot {\rm e}')  <0, 
\label{recausal} 
\end{equation}
and yields the corresponding global causal ordering on 
$\widehat {\cal C}_{2,d}$   
\begin{equation}
(\tau,{\rm e}) > (\tau',{\rm e}')\ \ \ {\rm  iff}\  \ \ \tau-\tau' > 
2 {\mathrm {Arcsin}} \left(\frac{({\rm e}-{\rm e}')^2}{4}
 \right)^{\frac{1}{2}}, 
\label{retimeorder}
\end{equation}
equivalently written e.g. in \cite{Luscher:1975ez}  as  
$ 
\tau-\tau' > 
{\mathrm {Arccos}} {({\rm e}\cdot {\rm e}')}. $
Note that in the space of variables $(\tau,\tau',{\rm e},{\rm e}')$, 
the region described by Eq.(\ref{retimeorder}) is 
exactly the limit of the region given by Eq.(\ref{timeorder}) 
when $r$ tends to infinity.
 
By taking the intersection of 
${\cal C}_{2,d}$ with the family of hyperplanes with equation $\eta^d + 
\eta^{d+1} = e^v$, 
one obtains the analogue of the horocyclic parametrization (\ref{coordinates}), namely:
\begin{equation}
\left\{\begin{tabular}{lclcll}
 $\eta^{\mu} $&=& $e^{ {v}} x^\mu  $ & ${ \mu=0,1,...,d-1}$\cr
 $\eta^{d} $&=& $ \frac 12 e^v (1+ x^2) $   
 &\ \  $x^2 ={ x^0}^2- {x^1}^2- \cdots -{x^{d-1}}^2$  \cr
 $\eta^{d+1}$&=&$\frac 12 e^v (1-  x^2) $  
\label{horocoordinates}
\end{tabular}\right. 
\end{equation}
which implies the following identity (similar to (\ref{7})) between quadratic forms:
\begin{equation}
(\eta -\eta')^2 = e^{v+v'} (x-x')^2
\label{horocausal}
\end{equation}

By taking Eqs. (\ref{conecoordinates}) into account,
one then sees that these formulae correspond (in dimension $d$) 
to the embedding of Minkowski space  
into the covering of the cone ${\cal C}_{2,d}$ (see \cite{Todorov} and references therein), namely 
one has (in view of the identification $ \eta^d +\eta^{d+1} = e^v = r( {\rm
e}^d + \cos \tau)$):  
\begin{equation}
x^0 = \frac {\sin \tau}{\cos \tau + {\rm e}^d},\ \ \ x^i = \frac {{\rm e}^i}
{\cos \tau + {\rm e}^d}, 
\label{embed}
\end{equation}
with
\begin{equation}
\cos \tau +{\rm e}^d >0,\ \  -\pi < \tau < \pi.
\end{equation}

\vskip 0.4cm
Let us now consider a general QFT on
$\widehat {AdS}_{d+1}$ whose Wightman functions ${\cal W}_n$ satisfy AdS invariance
together with the properties  a),b) and c) described in the previous section. 
In view of c), we can associate with the latter the following set of $n-$point 
distributions 
$\widetilde {\cal W}_n (\eta_1,...,\eta_n)$ on  
$\widehat {\cal C}_{2,d}$:   

\begin{equation}
\widetilde {\cal W}_n (\eta_1,...,\eta_n) =
 (r_1 \cdots r_n)^{-\Delta}
{\cal W}_n^{\infty}([\tau_1,{\rm e}_1],...,[\tau_n,{\rm e}_n]).  
\label{19}
\end{equation}
At first, one can check that the set of distributions  
$\widetilde {\cal W}_n $ satisfy the required 
positivity conditions for defining
a QFT on
$\widehat {\cal C}_{2,d}$.   
This is because, in view of postulate c) (applied with all $r_j$ equal to
the same $r$), the distributions ${\cal W}_n^{\infty}$ 
appear as the limits of the $n$-point functions of the QFT's  
on the spacetimes $\hat C_r$ when $r$ tends to infinity. 
The positivity conditions satisfied by the latter are then preserved in the limit,
in terms of test-functions  
of the variables $\tau_j$ and ${\rm e}_j$, and then extended in a trivial
way into the radial variables $r_j$ as positivity conditions 
for the distributions on the cone  
$\widehat {\cal C}_{2,d}$ (by using the appropriate test-functions 
homogeneous in the variables $r_j$ \cite{Luscher:1975ez}).  
\vskip 0.4cm
It follows from the reconstruction procedure \cite{Streater} that the set of distributions  
$\widetilde {\cal W}_n $ define a quantum field $\tilde O (\eta)$ on  
$\widehat {\cal C}_{2,d}$.
$\tilde O (\eta)$ enjoys the following properties:\\[10pt] 
{\em Local commutativity:}
Since the region (\ref{retimeorder}) is the limit of (\ref{timeorder}) for $r$ tending to infinity, it
results from the boundary condition c) and from the local commutativity of all fields $\Phi_r$ 
in the corresponding spacetimes $\hat C_r$ that the field $\tilde O (\eta)$
satisfies local
commutativity on $\widehat {\cal C}_{2,d}$.
\vskip 0.4cm
\noindent
{\em Spectral condition:}
In view of our postulate c) extended to the complex domain $T_n$ in the variables $\tau$, 
we see that the $n-$point distributions $\tilde {\cal W}_n (\eta_1,...,\eta_n)$ are boundary values 
of holomorphic functions in the same analyticity domains 
of $\left(\widehat {\cal C}_{2,d}^{(c)}\right)^n $  
as those of the L\"uscher-Mack field
theories \cite{Luscher:1975ez}. In particular, the restrictions of these holomorphic 
functions to the Euclidean space
domains $\{\eta= (\eta_1,...,\eta_n);\  \eta_j^0 = ir \sinh \sigma_j,\ \
\eta_j^i = r{\rm e}_j^i, \  
i=1,...,d,\ \ \eta_j^{d+1}
 = r \cosh \sigma_j;\ \  \sigma_1 <\sigma_2 <\cdots <\sigma_n\}$ 
yield the Schwinger functions of the theory.
It is also clear that, if the original Schwinger functions on the
complexified AdS space satisfy the cluster property (\ref{cluster}),
the corresponding Schwinger functions on $\widehat {\cal C}_{2,d}^{(c)}
$ satisfy the Luscher-Mack conformal cluster property (formula (5.1) of
\cite{Luscher:1975ez}) ensuring the uniqueness of the vacuum.

\vskip 0.4cm
We are now going to establish that the $\hat G-$invariance (\ref{covariance})
of the AdS $n-$point functions, together with the properties a), b), c), 
imply the conformal invariance of the field $\tilde O  (\eta)$; more 
precisely, we wish to show that the  Wightman functions    
$\widetilde {\cal W}_n $ of this field  
are invariant under the action 
on $\widehat {\cal C}_{2,d}$   
of the group $\hat G$, now interpreted as in \cite{Luscher:1975ez}  
as the ``quantum mechanical conformal group'', namely that one has: 

\begin{equation}
\widetilde {\cal W}_n (g\eta_1,...,g\eta_n) =   
\widetilde {\cal W}_n (\eta_1,...,\eta_n)   
\label{qmcf}
\end{equation}
for all $g$ in $\hat G$.

A part of this invariance is trivial in view of the limiting procedure of c): it is 
the invariance under the rotations in the $(0,d+1)-$plane (i.e. the 
translations in the time variables $\tau$) 
and the invariance under the spatial orthogonal group of the subspace of variables $ (\eta^1,...,\eta^d)$  
(acting on the sphere ${\mathbb S}_{d-1}$).

In order to show that the invariance condition (\ref{qmcf}) holds for all $g$ in $\hat G$, it 
remains to show that it holds for all one-parameter subgroups of
pseudo-rotations in the $(0,i)-$planes and 
in the $(i,d+1)-$planes of coordinates, with $i=1,...,d$. 
Let us consider the first case with e.g. $i=1$ and associate with the  
corresponding subgroup $G_{0,1}$ of pseudo-rotations the following parametrizations 
$X=X\{\rho,\psi,u\}$ and $\eta =\eta\{\rho,\psi,u\}$ (with $u=
(u^2,...,u^{d+1})$)  
of $\widehat {AdS}_{d+1}$ 
and of $\widehat {\cal C}_{2,d}$:
\begin{eqnarray}
&&\left\{\begin{tabular}{lclcll}
$X^{0} $  &=& $\rho \sinh \psi $ & \cr
$X^{1} $  &=& $\rho \cosh \psi $ & \cr                  
$X^{i}$ &=&$ \sqrt {{\rho}^2 +1} \   u^i$& 
& ${ i=2,...,d}$ & \cr 
$X^{d+1}$ &=&$ \sqrt {{\rho}^2 +1}\   u^{d+1} $& 
&${u^{d+1}}^2- {u^2}^2 - \cdots -{u^d}^2 =1$ & 
\label{pseudoads} 
\end{tabular}\right.
\\
&& \left\{\begin{tabular}{lclcll}
$\eta^{0} $  &=& $\rho \sinh \psi $ & \cr
$\eta^{1} $  &=& $\rho \cosh \psi $ & \cr
$\eta^{i} $  &=& $\rho\   u^i $ &                 
& ${ i=2,...,d}$ & \cr 
$\eta^{d+1}$ &=&$\rho\  u^{d+1} $& 
&${u^{d+1}}^2- {u^2}^2 - \cdots -{u^d}^2 =1$ & 
\label{pseudocone} 
\end{tabular}\right.
\end{eqnarray}
For $g \in G_{0,1},$ the invariance condition (\ref{qmcf}) to be proven can be written as follows
(with the simplified notation   
$\widetilde {\cal W}_n (\eta_1,...,\eta_n) =   
\widetilde {\cal W}_n (\eta_j)$):
\begin{equation}
\widetilde {\cal W}_n (\eta_j\{\rho_j,\psi_j +a,u_j\})  
= \widetilde {\cal W}_n (\eta_j\{\rho_j,\psi_j,u_j\})   
\label{psi-invar}
\end{equation}
for all real $a$.  
Now in view of the definition (\ref{19}) of  
$\widetilde {\cal W}_n (\eta_j)$ and of the relations between the sets of
parameters $(r,\tau,{\rm e})$  
and $(\rho,\psi,u)$ obtained by identification of the expressions 
(\ref{conecoordinates}) and (\ref{pseudocone}) of $\eta$, the invariance
condition (\ref{psi-invar})
to be proven is equivalent to the following condition for the
asymptotic forms of the AdS $n-$point functions 
${\cal W}_n^{\infty}$ (for all $a$): 
\begin{eqnarray}
&&
\prod_{1 \leq k \leq n} 
\left( (\sinh \psi_k)^2 +(u_k^{d+1})^2 \right)^{-\frac{\Delta}{2}}  
\left( (\sinh (\psi_k+a))^2 +(u_k^{d+1})^2 \right)^{\frac{\Delta}{2}}\times \cr
&&
{\cal W}_n^{\infty}
\left(\left[ {\mathrm {arctg}} \frac{\sinh \psi_j}{u_j^{d+1}},
\frac{\cosh \psi_j} 
{ \left( (\sinh \psi_j)^2 +(u_j^{d+1})^2 \right)^{\frac{1}{2}}},
\frac {u_j^i}
{ \left( (\sinh \psi_j)^2 +(u_j^{d+1})^2 \right)^{\frac{1}{2}}}
 \right]\right)= \cr
&&  
{\cal W}_n^{\infty}\left(\left[{\mathrm  {arctg}} \frac{\sinh (\psi_j +a)}{u_j^{d+1}},
\frac{\cosh \psi_j} 
{ \left( 
(\sinh (\psi_j +a))^2 +(u_j^{d+1})^2 \right)^{\frac{1}{2}}},\right.\right.
\cr
&&
\qquad\qquad\qquad\qquad\qquad\qquad\qquad \left.\left.\frac {u_j^i}
{ \left( (\sinh (\psi_j +a))^2 +(u_j^{d+1})^2 \right)^{\frac{1}{2}}}
\right]\right) .  
\label{30}
\end{eqnarray} 
In this equation the 
symbol arctg($\cdot$)  denotes the angle $\tau_j$ of the parametrization 
(\ref{conecoordinates}),
 which can take all real values; however, one notices that under the 
transformation $\psi_j \to \psi_j +a$, 
the angle $\tau_j$ varies in such a way that the point 
$\eta$ remains in the same sheet of the covering 
$\widehat{\cal C}_{2,d}$ of the cone    
${\cal C}_{2,d}$ (e.g. $-\frac{\pi}{2} 
< \tau_j < \frac{\pi}{2}$ for the choice of Arctg).      

Comparing the parametrizations (\ref{sphericcoordinates}) and
(\ref{pseudoads})  
of $\widehat {AdS}_{d+1}$ we obtain the following relations:
\begin{eqnarray}  
&& r=\rho \left[  (\sinh \psi)^2 + (u^{d+1})^2   + 
\frac {(u^{d+1})^2 - 1}{\rho^2} 
\right]^{\frac{1}{2}} \equiv \rho \ h_{\rho}(\psi,u^{d+1}) 
\label{31}
\\
&&  
\tau = {\mathrm {arctg}} \left[ \left(1 +\frac{1}{\rho^2}\right)^{-\frac{1}{2}} 
\frac {\sinh\psi}{u^{d+1}}\right]   
\label{32}
\\
&&  
{\rm e}^1 =\frac{\cosh \psi}{h_{\rho}(\psi,u^{d+1})},\ \ 
{\rm e}^i =\frac{u^i}{h_{\rho}(\psi,u^{d+1})}\left(1 + \frac{1}{\rho^2}\right)^{\frac{1}{2}}.  
\label{33}
\end{eqnarray} 
Note that  the 
function $h_{\rho}$ introduced in (\ref{31}) is such that
\begin{equation}
\lim_{\rho \to \infty}  h_{\rho}(\psi,u^{d+1}) =  
 \left[  (\sinh \psi)^2 + (u^{d+1})^2    
\right]^{\frac{1}{2}} 
\end{equation}
This implies that it is equivalent   
to take the limits in Eq. (\ref{15}) 
for $\rho_j$ (instead of $r_j$) tending to infinity and at fixed value of
$\psi_j$ and $u_j$, after plugging  the expressions (\ref{31}), (\ref{32}),
(\ref{33}) of $r_j, \tau_j, {\rm e}_j$ into both sides of Eq. (\ref{15}):
$$\lim_{min(\rho_1,...,\rho_n) \to {+\infty}}
\left|(\rho_1\cdots \rho_n)^{\Delta} {\cal W}_n
(X_j\{\rho_j,\psi_j,u_j\}) 
-\prod_{1 \leq k \leq n} h_{\rho_k}(\psi_k,u_k^{d+1})^{-\Delta}\ \right. \times $$ 
\begin{equation}  
\times\left.{\cal W}_n^{\infty}\left(\left[
{\mathrm {arctg}} \left[   
\frac {\sinh\psi_j}
{\left(1 +\frac{1}{\rho_j^2}\right)^{\frac{1}{2}}u_j^{d+1}}\right],   
\frac{\cosh \psi_j}{h_{\rho_j}(\psi_j,u_j^{d+1})}, 
\frac{u_j^i \left(1 + \frac{1}
{\rho_j^2}\right)^{\frac{1}{2}}}{h_{\rho_j}(\psi_j,u_j^{d+1})}  
\right]\right)\right|  = 0. 
\label{34}
\end{equation} 
If we now also
 consider the vanishing limit of the same difference after the transformation
$\psi_j \to \psi_j +a$ has been applied, and take into account the fact that,
by assumption, 
the first term of this difference has remained unchanged, we obtain the following  
relation:
$$\lim_{min(\rho_1,...,\rho_n) \to {+\infty}}
\left|\prod_{1 \leq k
 \leq n} h_{\rho_k}(\psi_k,u_k^{d+1})^{-\Delta}\ \right. \times $$ 
$$  
\times\left.{\cal W}_n^{\infty}\left(\left[
{\mathrm {arctg}} \left[   
\frac {\sinh\psi_j}
{\left(1 +\frac{1}{\rho_j^2}\right)^{\frac{1}{2}}u_j^{d+1}}\right],   
\frac{\cosh \psi_j}{h_{\rho_j}(\psi_j,u_j^{d+1})}, 
\frac{u_j^i \left(1 + \frac{1}
{\rho_j^2}\right)^{\frac{1}{2}}}{h_{\rho_j}(\psi_j,u_j^{d+1})}  
\right]\right)\right.
$$
$$-\prod_{1 \leq k \leq n} h_{\rho_j}
(\psi_k  +a,u_k^{d+1})^{-\Delta}\  \times$$
 \begin{equation}  
\times\left.{\cal W}_n^{\infty}\left(\left[
{\mathrm {arctg}} \left[   
\frac {\sinh(\psi_j +a)}
{\left(1 +\frac{1}{\rho_j^2}\right)^{\frac{1}{2}}u_j^{d+1}}\right],   
\frac{(\cosh \psi_j+a)}{h_{\rho_j}(\psi_j+a,u_j^{d+1})}, 
\frac{u_j^i \left(1 + \frac{1}
{\rho_j^2}\right)^{\frac{1}{2}}}{h_{\rho_j}(\psi_j+a,u_j^{d+1})}  
\right]\right)\right|
 = 0.  
\end{equation} 
Now it is easily seen
 that in the latter, the limit can be taken separately in each term
and that the resulting equality yields precisely 
the required covariance relation (\ref{30}) for 
${\cal W}_n^{\infty}$.

Although the previous formulae have been
written in terms
 of the distributions ${\cal W}_n$ and of their asymptotic forms, one could
reproduce the argument
 in a completely rigorous way \cite{Bros:1999} 
in terms of the functions $W_n$ in the tube 
domains $T_n$ of the
 variables $\tau_j$, all the functions involved being then of class
${\cal C}^{\infty}$ with respect to all the variables $(\rho_j,\psi_j,u_j)$  
and all the limits being taken in the sense of 
regular functions; the covariance 
relations on the reals will then be obtained  as relations for the
corresponding boundary  values
(in the sense of distributions).   
The treatment of the covariance
 with respect to the pseudo-rotation groups $G_{i,d+1}$ is completely
similar. 
\vskip 0.4cm
\noindent
We can then summarize the results of  this section by the following statement:
\vskip 0.4cm
\noindent
{\em the   procedure we have described 
(expressed by Eqs. (\ref{15}) and (\ref{19}))  
displays a general AdS/CFT correspondence for QFT's: 
\begin{equation} \Phi(X)
\rightarrow \tilde{\mathcal O}(\eta)\   
\label{correspondence}
\end{equation}
between a scalar (AdS invariant) quantum field $\Phi(X)$ on the covering 
$\widehat {AdS}_{d+1}$ of  
$ {AdS}_{d+1}$ whose Wightman functions satisfy the  properties 
a),b),c), 
and a conformally invariant local field $\tilde{\mathcal O}(\eta)$   
on the covering 
$\widehat{\cal C}_{2,d}$ of the cone    
${\cal C}_{2,d}$, enjoying     
the L\"uscher-Mack spectral condition; the degree of homogeneity (dimension)  $\Delta$ of $\tilde{\mathcal O}(\eta)$ is equal to the asymptotic dimension  
of the AdS field $\Phi(X)$.} 
\vskip 0.4cm
Of course, from this general point of view, the correspondence may a priori 
 be many-to-one.
Finally, according to the formalism described in \cite{Luscher:1975ez,Todorov}, the correspondence 
(\ref{correspondence}) can be completed by saying that there exists a unique  
conformal (Minkowskian) local field ${\mathcal O}(x)$ of dimension $\Delta$ 
whose $n-$point functions ${\cal W}_n^M$ are expressed in terms of those of 
$\tilde{\mathcal O}(\eta)$ by the following formulae:   
\begin{equation}  
{\cal W}_n^M(x_1,...,x_n) =   
e^{(v_1+\cdots+v_n)\Delta}\  \widetilde {\cal W}_n (\eta_1,...,\eta_n) =  
\Pi_{1 \leq j \leq n} (\eta_j^d + \eta_j^{d+1})^{\Delta}\  \widetilde {\cal W}_n (\eta_1,...,\eta_n) .  
\label{Mink}
\end{equation} 
In the latter, the Minkowskian variables $x_j$ are expressed in terms of the cone variables 
$\eta_j$ by inverting (\ref{horocoordinates}), which yields:
\begin{equation}  
x_j^{\mu} = \frac{\eta_j^{\mu}}{\eta_j^d + \eta_j^{d+1}}. 
\label{invertMink}
\end{equation}

\section{Two-point functions}

\subsection{The analytic structure of two-point functions on the AdS spacetime}

It turns out that in all field theories 
on $\widehat {AdS}_{d+1}$
satisfying  
the general requirements described in subsection 2.2, 
the two-point function  
enjoys {\em maximal analyticity properties} 
in all the  coordinates, as it is the case for the Minkowski \cite{Streater} 
and de Sitter cases \cite{Bros:1996js}. 
A full proof of these results  will be found in \cite{Bros:1999}. 
We shall only
 give here a descriptive account of them, needed for further applications.
Since, in particular, AdS
 covariance and the ``energy spectrum condition'' b) of Sec 2.2 
are responsible for
 this maximal analytic structure and since (as seen below) the latter determines 
completely  satisfactory solutions for the case of Klein-Gordon AdS fields, 
we shall consider this general class of two-point functions as   ``preferred''.

There are two distinguished complex domains \cite{Bros:1999} of   
$AdS^{(c)}_{d+1}$, invariant
 under real AdS transformations, which are of crucial importance for a full
understanding of the structures associated with two-point functions.
They are given by:
$$ T^+  =\{ Z  = X+iY  \in AdS^{(c)}_{d+1}; \, Y^2>0,\, \epsilon(Z) = +1\},$$ 
\begin{equation}
T^-  =\{ Z = X+iY  \in AdS^{(c)}_{d+1}; \, Y^2>0,\, \epsilon(Z) = -1\}, 
\label{Tubes}
\end{equation}
where 
\begin{equation}
\epsilon(Z) =  \mbox{sign}  (Y^0 X^{d+1}- X^0 Y^{d+1}).
\end{equation}
$T^+$ and $T^-$  are the AdS version of 
the usual forward and backward tubes $T_{  M}^+$ and $ T_{  M}^-$ 
of complex Minkowski spacetime, obtained in correspondence with 
the energy-momentum spectrum condition \cite{Streater};
let us recall their  definition 
(in arbitrary spacetime dimension $p$):
 \begin{eqnarray}
& T_{  M}^+ & =\{ z = x+i y   \in {{\mathbb M}^{p}}^{(c)}; \, y^2>0,\, y^0>0\}, \cr
&   T_{ M}^- & =\{ z = x+i y   \in {{\mathbb M}^{p}}^{(c)}; \, y^2>0,\, y^0<0\}. 
\label{tubesminkowski}
\end{eqnarray}
In the same way as these Minkowskian tubes are generated by the action of 
real Lorentz transformations on the ``flat'' (one complex time-variable) domains 
$\{z = x+iy;\  y=(y^0,\vec 0);\  y^0 >0 \ ({\makebox{resp. }}  y^0 <0)\}$, 
the domains (\ref{Tubes}) of  
$AdS^{(c)}_{d+1}$  are generated by the action of the group $G$  
on the flat domains obtained by letting $\tau$ vary in the half-planes $\Im \tau >0$ or 
$\Im \tau <0$ and keeping $r$ and $e$ real in the 
covering parametrization (\ref{sphericcoordinates}) of the AdS quadric.
In fact, by using the complex extension of this parametrization and putting 
$r= \sinh (\psi +i \phi)$, $\tau = \Re \tau + i \sigma $ one can represent  
the domains (\ref{Tubes}) by the following semi-tubes (invariant under translations 
in the variable $\Re \tau$): 
\begin{equation}
\pm \sinh \sigma >  \left[\frac{(\sin \phi)^2 + \left( (\cosh \psi)^2 - (\cos \phi)^2 \right)
(\Im e)^2}
{  (\cosh \psi)^2 - (\sin \phi)^2}\right]^\frac{1}{2} 
\label{retubes}
\end{equation}
This representation (which clearly contains the previously mentioned flat domains) 
can be thought of, either as representing the domains  
(\ref{Tubes}) of $AdS^{(c)}$  
if $\tau$ is identified to $\tau + 2 \pi$, or
coverings of the latter embedded in 
$\widehat{AdS}^{(c)}_{d+1}$, which we denote by  
$\hat T^+$ and 
$\hat T^-$, if one does not make this identification.

One typical property of Wightman's QFT \cite{Streater}
is that any two-point distribution  ${\mathcal W}_M(x,x')$  satisfying 
the spectral condition 
is the boundary value of a  function $W_M(z,z')$ holomorphic for $z\in T_{ M}^-$
and $z'\in T_{M}^+$. An analogous property also holds for 
$n$-point functions.

It is a consequence of AdS invariance together with the spectrum assumption b) \cite{Bros:1999} 
that, also 
in the AdS spacetime,  general 
two-point functions  can be  
characterized by the following global analyticity property which plays the role of a  
{\em $G-$invariant spectral condition}: \\[5pt]
b$^{({\mathrm {inv}})})$  {\em Normal analyticity condition for two-point functions:} 
{\em the two-point function $\mathcal W\left( X, X'\right)$
is the boundary value of a  function $ W\left(
Z, Z'\right) $ which is holomorphic in the  domain 
$\hat{T}^-\times \hat{T}^+$ of 
$\widehat{AdS}^{(c)}_{d+1}\times \widehat{AdS}^{(c)}_{d+1}$.}
\\[5pt]
A further use of AdS invariance implies that 
$W(Z,Z')$ is actually a function $w(\zeta)$ 
of a single complex variable $\zeta$;  
this variable $\zeta$ can be identified with 
$Z\cdot Z'$ when  $Z$ and $Z'$ are both in the fundamental  sheet of $\widehat{AdS}^{(c)}_{d+1}$;  
AdS invariance and the normal analyticity condition together imply the following \\[10pt]
{\bf  Maximal analyticity property}:
{\em $w(\zeta)$  
is analytic in the  covering $\widehat\Theta$ of the
cut-plane $\Theta = \{{\mathbb C} \setminus[-1,1]\}$.} \\[10pt] 
 For special theories which are periodic in the time coordinate
$\tau$, $w(\zeta)$ is in fact analytic in $\Theta$ itself.  
One can now introduce all the usual Green functions.
The ``permuted Wightman function'' ${\mathcal W}(X',X) = 
\langle \Omega, \Phi(X') \Phi(X) \Omega \rangle$ 
is the boundary value of 
${ W}(Z,Z')$ from the domain 
$\{(Z,Z'): Z\in { \hat  T}^{+}, \; Z'\in {\hat  T}^{-}\}$.
The commutator function is then  
$\mathcal C(X,X')={\mathcal W}(X,X')-{\mathcal W}(X',X) $.
The retarded propagator ${\mathcal R}(X,X')$ is introduced by 
splitting the support
of the commutator ${\mathcal C}(X,X')$ as follows 
\begin{equation}
{\mathcal R}(X,X')= i\theta (\tau - \tau') { \mathcal C}(X,X'),
\end{equation}
The other Green functions  are then defined in terms of 
${\mathcal R}$  by the usual formulae:
the advanced propagator is given by ${\mathcal A}= { \mathcal R}-i{ \mathcal C}$ while the 
chronological propagator is  given by 
${\mathcal F}= -i{\mathcal A}+{\mathcal W}$.  

Note finally that, as a function of the single variable $\zeta = X\cdot X'$, the jump 
$i\delta w(\zeta)$ of 
$i w(\zeta)$  across its cut $(-\infty,+1]$ coincides with  the retarded  
propagator $R(X,X')$ (or the advanced one); in the periodic (i.e. ``true AdS'') case, the support of 
$\delta w$ reduces to the compact interval $[-1,+1]$.  

\subsection{The simplest example revisited:
 Klein-Gordon fields in the AdS/CFT correspondence}

The Wightman functions of fields satisfying the  Klein-Gordon equation  $AdS_{d+1}$
\begin{equation}
\square_{AdS} \Phi + m^2 \Phi = 0.
\label{kg}
\end{equation}
display the simplest example of the previous analytic  structure:
\begin{equation}
W_{\nu}(Z,Z') = w_\nu(\zeta) = 
\frac {e^{-i\pi\frac {d-1}2}}{(2\pi)^{\frac{d+1}2}} 
(\zeta^2-1)^{-\frac {d-1}4} Q^{\frac {d-1}2}_{\nu-\frac 1 2}(\zeta).
\label{kgtp}
\end{equation}
Here $Q$ is a second-kind Legendre's 
function\footnote{This is the way these Wightman functions were first written 
in \cite{Fronsdal:1974} for the four dimensional case $d=3$. 
Their identification with second--kind Legendre functions is worth being
emphasized, in place of their less specific (although exact) introduction
under the general label of hypergeometric functions, used in recent papers.
In fact Legendre functions are basically linked to the geometry of the dS and
AdS quadrics from both group--theoretical and complex analysis viewpoints
\cite{Bros:1996bv2, Bros:1996js, faraut, vilenkin}
}
\cite{Bateman};
the parameter $\nu$  is linked to the field's mass by the relation 
\begin{equation}
\nu^2  = \frac {d^2} {4} + m^2.
\label{nu}
\end{equation}
and the normalization of $W_\nu$ is chosen by imposing 
the short-distance Hadamard behavior. 

Since $W_{\nu}(Z,Z')$ and  $W_{-\nu}(Z,Z')$ are solutions of 
the same Klein-Gordon equation (and share the same analyticity properties), 
the question arises if these Wightman function both define
acceptable QFT's on $AdS_{d+1}$.
The answer \cite{Breitenlohner:1982jf} is that only 
theories with $\nu \geq -1$ 
are acceptable and there are therefore two regimes: 
for $\nu > 1$ there is only one field theory corresponding to a given mass
while for $|\nu| < 1$ there are two theories. The case $\nu = 1$ is a limit case.
Eq. (\ref{kgtp}) shows clearly that
the only difference between the theories parametrized by opposite values of $\nu$ is in their
large distance behavior. More precisely, in view of Eq. (3.3.1.4) of \cite{Bateman}, we can write:  
\begin{equation}
w_{-\nu} (\zeta)  = w_{\nu} (\zeta)  + \frac{\sin \pi\nu}{(2\pi)^{\frac{d+1}{2}}}
\, \Gamma\le(\frac d 2-\nu\ri)
\Gamma \le(\frac d 2 +\nu\ri) (\zeta^2-1)^{-\frac {d-1}4}P^{-\frac {d-1}2}_{-\frac 1 2 -\nu}(\zeta) .
\end{equation}
Now we notice that in this relation (where all terms are solutions of the same Klein-Gordon equation)   
the last term is {\em regular 
on the cut}  $\zeta \in [-1,1]$. 
This entails (reintroducing the AdS radius $R$) that, in the two theories, the
$c-$number commutator 
$[\Phi(X),\Phi(X')]$ takes the same value for all (time-like separated)
vectors $(X,X')$ such that  
$ |X\cdot X'| < R^2 $.  
Therefore we can say that {\em the two theories represent the same algebra of
local observables   
at short distances (with respect to the radius $R$)}. 
But since the last term in the latter relation grows the faster the 
larger is $|\nu|$ (see \cite{Bateman} Eqs. (3.9.2)), we see that  
the two theories drastically differ by their long range behaviors. 

The  existence of the  two regimes above  has given rise
to two distinct treatments of the AdS/CFT correspondence in the two cases
\cite{Klebanov:1999tb} 
and symmetry breaking had been advocated to explain the difference.

In the present context, by applying the correspondence as given in Eq. 
(\ref{correspondence}), the two regimes can be  treated in one stroke.
Indeed, 
Eq. (3.9.2.21) of \cite{Bateman} reports the following
large $\zeta$ behavior of  the Legendre's function $Q$ (valid for any complex $\nu$):
\begin{equation}
Q^{\frac {d-1}2}_{\nu - \frac 1 2 } (\zeta) \simeq e^{i\pi \frac {d-1}2
} 2^{-\nu-\frac 12 } \frac{\Gamma\le(\nu+\frac d 2
\ri)}{\Gamma\le(\nu+1\ri)} \pi^\frac 12 \zeta^{-\frac 1 2  -\nu} \ .
\end{equation}
It follows that the two-point function (\ref{kgtp}) and thereby all 
the $n-$point functions of the corresponding 
Klein-Gordon field satisfy the 
dimensional boundary conditions at infinity 
with dimension $\Delta = \frac{d}{2} + \nu$. Indeed, let $\tau$ and $\tau'$ 
be complex and such that $\Im \tau < \Im \tau'$. It follows that 
\begin{eqnarray}
{W}_\nu^{\infty}([\tau,{\rm e}],[\tau',{\rm e}']) & = &\lim_{r, r' \to \infty} 
{(rr')}^{\frac{d}{2} + \nu}W_{\nu}(Z[\tau,r,{\rm e}],Z'[\tau',r',{\rm e}']) =\cr
&=& \frac{2^{-\nu -1}}{(2\pi)^\frac{d}{2}} 
 \frac 
{\Gamma(\nu+ \frac {d}2 )}{\Gamma(\nu+1)}
\frac1{{[\cos(\tau - \tau') - {\rm e}\cdot {\rm e}']}^{\frac {d}2+\nu}}.
\label{limit1} 
\end{eqnarray}
(see also \cite{Giddings:1999jq}).  
This equation expresses  nothing more  
than the  behavior of
the previous Legendre's function at infinity.
Not only all the $\nu$'s are treated this way
in one stroke but, also,   one can study   the boundary 
limit  for theories corresponding to $\nu<-1$, 
even if the corresponding QFT may have no direct physical interpretation. 

The two-point function of the  conformal field 
$\tilde{ \mathcal O}(\eta)$ on the cone $\widehat {\cal C}_{2,d}$ 
corresponding to (\ref{limit1}) is then constructed   
by following the prescription of Eq.(\ref{19}),  
which yields 
\begin{equation}
  \widetilde { W}_{\nu}  (\eta ,\eta') =
  (r r')^{-\frac{d}{2} - \nu}
{ W}_{\nu}^{\infty}([\tau,{\rm e}],[\tau',{\rm e}']) =  
 \frac{1}{2\pi^\frac{d}{2}} 
 \frac 
{\Gamma(\nu+ \frac {d}2 )}{\Gamma(\nu+1)}
\frac{1}{{[-(\eta -\eta')^2]}^{\frac {d}{2}+\nu}}.
\label{limit2} 
\end{equation}
Correspondingly, we can deduce from (\ref{limit2}) 
the expression of the two-point function of the associated Minkowskian
field 
on $\mathbb M^d$, given by formula (\ref{Mink});  
by taking Eq.(\ref{horocausal}) into account,  
we obtain:
\begin{equation}
 W_{\nu}^M(z,z') =  e^{(v+v')
(\frac{d}{2} +\nu)}\widetilde { W}_{\nu} \left( \eta\left(v,z \right) ,
\eta'\left(v',z'\right)\right)= 
\frac { 1 }{2\pi^{\frac{d}2}}  \frac 
{\Gamma(\nu+ \frac {d}2 )}{\Gamma(\nu+1)} 
\frac1{{[-(z-z')^2  ]}^{\frac {d}2+\nu}}.
\label{gurru}
\end{equation}
In the latter, 
the Poincar\'e coordinates 
$z$ and $z'$ must be taken with the usual $i\epsilon-$prescription  
($\Im z^0 < \Im z'^0$),  
which can be checked to be implied by the spectral condition b) of section 2
through the previous limiting procedure. \par \vskip 5pt

We note that this natural way of producing the boundary
field theory gives rise to
the normalization advocated in
\cite{Klebanov:1999tb}, Eq.  (2.21) (apart from a trivial 
factor 4, which does not depend
on the anomalous dimension $\Delta= \frac{d}{2} + \nu$).\\

Let us now describe how the previous limiting procedure looks in the
Poincar\'e\ coordinates (\ref{coordinates}).   These coordinates offer
the possibility of studying directly the boundary behavior of the AdS Wightman
functions in a larger domain of the complex AdS spacetime.  This fact
is based on the following simple observation:  consider the  parametrization
(\ref{coordinates}) for two points with complex parameters specified by
\begin{eqnarray}
 & Z= Z( {v}, z), \qquad      &{v} \in \mathbb R, \; z \in { T_M ^-} \cr
& Z'= Z'( {v'}, z'), \qquad &{v'} \in \mathbb R, \; z' \in { T_M ^+}. 
\end{eqnarray}
It is easy to check that this choice of parameters 
implies that $Z \in {T ^-}$ and  $Z' \in {T ^+}$. 
It follows that, given an AdS invariant two-point function satisfying 
locality and the normal analyticity condition $b^{(inv)})$,  
the following restriction automatically generates
a local and (Poincar\'e) covariant two-point function on the slice $\Pi_v$,
which satisfies the spectral condition \cite{Streater} (in short: the two-point 
function of a general Wightman QFT):
\begin{equation}
{W} ^{M}_{\{ v\}}(z,  z')=
W \left(Z( {v},z), Z'( {v},z')\right).
\label{restrtp}
\end{equation}
On the basis of the dimensional boundary condition (\ref{15}),  
and of the fact (obtained by comparing (\ref{sphericcoordinates}) and
(\ref{coordinates})) 
that $\frac{e^v}{r} = \sqrt {1+ \frac{1}{r^2}} \cos \tau +{\rm e}^d$
tends to the finite
 limit $\cos \tau + {\rm e}^d$ when $r$ tends to infinity, one sees that  
the following limit exists
 and that it yields (in view of (\ref{19}) and (\ref{Mink})): 
\begin{equation}
\lim_{v \to {+\infty}}e^{2v\Delta} {W} ^{M}_{\{ v\}}(z,  z')  
 = { W}^{M}(z,z').  
\label{55}
\end{equation}
The limiting two-point function 
${ W}^{M}(z,z')$ then automatically  
exhibits locality, Poincar\'e invariance and the spectral condition.
(The invariance under special 
conformal transformations and scaling property would necessitate a special check,
but they result from  
the general statement of 
conformal invariance  of the limiting field $\tilde {\mathcal O}(\eta)$ 
proved in section 3 completed by the analysis of 
\cite{Luscher:1975ez}).

When applied to the Wightman functions of Klein-Gordon fields (i.e. with
$\Delta = \frac d 2 +\nu$), the latter
presentation  
of the limiting procedure 
gives immediately the result obtained in Eq.(\ref{gurru})
but in a larger complex domain: 

\begin{eqnarray}
\lim_{ {v}\to \infty} e^{2 {v}\left(\frac {d}2+\nu\right)}
W_{\nu} (Z( {v},z) ,Z'( {v},z')) &=& 
\frac { 1 }{2\pi^{\frac{d}2}}  \frac 
{\Gamma(\nu+ \frac {d}2 )}
{\Gamma(\nu+1)} \frac1{{[-(z-z')^2]}^{\frac {d}2+\nu}} \label{corrscalar}
\end{eqnarray}

In a completely similar way one can compute the bulk--to--boundary
correlation function by considering a two-slice restriction $W_{\nu} (Z
( {v},z) ,Z'( {v'},z'))$
of $W_{\nu}$. The bulk-to-boundary correlation function is obtained 
by sending 
${v}'\to \infty$ while keeping
$ {v} $ fixed, by the following limit: 
\begin{eqnarray}
 \lim_{ {v'}\to \infty} e^{ {v}'\left(\frac {d}2+\nu\right)}
W_{\nu} (Z( {v},z) ,Z'( {v'},z')) &=&
 \frac 1{2 \pi^{\frac d 2 } }
\frac {\Gamma\le(\nu + \frac d 2 \ri) }{\Gamma(\nu +1 )} \frac 1
{\le(e^{- {v}} -e^ {v} (z-z')^2 \ri)^{\frac d 2 +\nu}} = \cr
&& = 
 \frac 1{2 \pi^{\frac d 2 } }
\frac {\Gamma\le(\nu + \frac d 2 \ri) }{\Gamma(\nu +1 )} 
\left(\frac {u }{u^2 -  (z-z')^2 }
\right)^{\frac d 2 +\nu}\ . 
\end{eqnarray}

\section{Decomposition      
of AdS Klein-\-Gor\-don fields to the branes}

We will discuss in this section a decomposition of Klein-Gordon fields
associated with the Poincar\'e coordinate system (\ref{coordinates}).
This will produce some new and exact formulae which exhibit
how a field of a given mass on the ambient AdS spacetime 
is decomposed into elementary massive fields when restricted to the brane.
We will also gain 
insight about the two different AdS regimes depending on the values of
of the mass parameter $\nu$. We follow here a method already used in \cite{Bertola:1999ga}.
   
According to Eq. (\ref{restrtp}), we can obtain  by restriction
Poincar\'e invariant  QFT's  on the branes $\Pi_v$ of
$AdS_{d+1}$.  Of course the restricted theories are not conformal and
can become conformal only in the limit $v\to\infty$.

Let us study the case of Klein-Gordon fields.    
By using the coordinates    
(\ref{coordinates})    
the Klein-Gordon equation (\ref{kg})    
is separated into the following pair of equations:   
\begin{eqnarray}   
&& \square_M \phi+ \lambda \phi = 0,  \\   
&& e^{2 {v}} \left[\theta''( {v}) + d \, \theta'( {v}) -m^2\theta( {v})\right] =   
-\lambda \theta( {v}).\label{equat}   
\end{eqnarray}    
The first equation is another Klein-Gordon equation, now considered    
on a $d$-di\-men\-sio\-nal Minkowski spacetime.   
The second equation is an eigenvalue equation for a second order   
operator. The separation constant $\lambda$ is for the moment unrestricted.   
To get information on the allowed values for $\lambda$ we have to   
consider Eq. (\ref{equat}) as   a spectral problem in a suitable    
 Hilbert space.   
To this end let us  introduce the Hilbert space    
$L^2( {\mathbb R} , e^{(d-2)v} {\mathrm d} v )$,    
where the differential operator defined in Eq. (\ref{equat}) is symmetric.   
It is useful to pass to the variable $u = e^{-v}$    
already introduced  in Eq. (\ref{coordinates}) and define    
$f(u)=\theta( {v}) e^{\frac{d-1}{2}  {v}} $. Eq. (\ref{equat})    
is then turned into       
\begin{equation}   
-f''(u) + \frac{m^2+\frac{d^2-1}4}{u^2} f(u)= -f''(u) +   
\frac{(\nu+1/2)(\nu-1/2)}{u^2} f(u)   
=\lambda f(u)\ ,   
\label{bessel}   
\end{equation}   
a well-known Schr\"odinger spectral problem on the half-line (the   
Hilbert space is now \hspace{-2pt}
 $L^2({\mathbb R} ^+\!\! ,\! {\rm d} u\!)\!$).   
   
Following \cite{Titchmarsh:1962}, pag. 88 ff, we learn that  there are   
two distinct regimes corresponding as before to $\nu \geq 1$ and     
$|\nu|<1$.    
   
When $\nu\geq 1$ the previous operator is essentially self--adjoint   
and  there is only one   
possible choice for the generalized eigenfunctions, namely   
 \be   
f_\l(u)=\frac 1 {\sqrt{2}} \,   u ^\frac 1 2 J_\nu\le(\sqrt{\lambda} \,u \ri)\ ,   
\ee   
where $J_\nu$ are Bessel's functions.   
The completeness of these eigenfunctions gives Hankel's formula,   
which expresses the   
resolution of the identity in $L^2({\mathbb R} ^+, {\rm d}u)$ as follows:   
\be   
g(u) = \int_{0}^\infty {\rm d}\l\, f_\l(u)\int_0^\infty f_\l   
(u') g(u') {\rm d} u'\ ,\ \ \forall g\in L^2({\mathbb R} ^+, {\rm d} u))\ .   
\ee   
When $0\leq \nu< 1$ both  solutions $u^{1/2}J_\nu(\sqrt{\lambda}u)$ and   
$u^{1/2}J_{-\nu}(\sqrt{\lambda} u) $  are square integrable in the
neighborhood of $u=0$ and    
must be taken into consideration:
we are in the so--called {\em limit circle case} at zero \cite{Titchmarsh:1962,Reed},   
which implies that the   
operator is not essentially self--adjoint and there exists a $S^1$    
ambiguity in the self--adjoint extensions we can perform.   
The freedom  is exactly in the choice of the boundary conditions at   
$u=0$ (corresponding to the   boundary of AdS).\\   
Now  we have a one--parameter family of eigenfunctions:   
 \be   
f^{(\K)}_\l (u) \equiv \sqrt{\frac {u} {2}}{{ \le(\K^2 - 2\K\l^\nu\cos(\pi\nu) +   
\l^{2\nu} \ri)^{-\frac 1 2}}}    
\le[\K\, J_{\nu}(\sqrt{\l}\, u) -\l^{\nu}  J_{-\nu}(\sqrt{\l}\, u) \ri]\    
,   
\ee    
to which we must add one  bound state when  $\K>0$:   
\be   
f^{(\K)}_{\rm bound}  (u) \equiv \sqrt{ 2\K^{\frac 1 \nu}\frac   
{\sin{\pi\nu}}{\pi\nu} } u^\frac 1 2  K_{\nu} (\K^{\frac 1 {2\nu}} u)\ .    
 \ee   
The possible choices of the parameter    
$\K$ do correspond to different self--adjoint extensions of the  differential   
operator (\ref{bessel}). To each such extension there is associated a   
domain ${\mathfrak D}^{(\K)}$ also depending on the parameter $\K$ \cite{Reed}.    
To construct ${\mathfrak D}^{(\K)}$ consider the one dimensional subspaces
$H_\pm$ spanned by the  eigenfunctions solving   
Eq. (\ref{bessel}) with eigenvalues $\pm i$:
\be   
f_{\pm} (u) \equiv \sqrt{u} K_\nu (e^{\pm \frac {i\pi}4} u)\ ;    
\ee   
both these functions are square-integrable when $0\leq \nu<1$.    
Each extension   
is in one--to--one correspondence with partial isometries   
$U:H_+\mapsto H_-$, namely --in this case-- with elements of   
$U(1)\simeq S^1$. The domain of the extension is obtained by   
adjoining to the original domain of symmetry the subspace $\le( {\rm   
id}_{H_+}+ U\ri) H_+$: here it means that we have to add the span of   
the $L^2$ element    
\bes   
f_\alpha(u)\equiv f_{+}(u) + e^{i\alpha} f_{-}(u)\ .   
\ees   
which has in our case the asymptotics   
\bea   
f_\alpha(u) \simeq \frac {\pi}{2\sin(\pi\nu)} \le[ \frac {2^\nu\le(   
e^{-\frac {i\pi\nu} 4} + e^{i\alpha + \frac {i\pi\nu}4}  \ri)}   
{\Gamma(1-\nu)} u^{-\nu} -  \frac {2^{-\nu}\le(   
e^{\frac {i\pi\nu} 4} + e^{i\alpha - \frac {i\pi\nu}4}  \ri)}   
{\Gamma(1+\nu)} u^{\nu}   
\ri]\ .   
\label{asymext}   
\eea   
The generalized eigenfunctions of the operator (\ref{bessel})   
corresponding to  a specific extension have the following asymptotics   
\be     
f^{(\K)}_\l(u) \simeq  2^{-\frac 1 2}   
  u^{\frac 1 2}  \le(\K^2 - 2\K\l^\nu\cos(\pi\nu) +   
\l^{2\nu} \ri)^{-\frac 1 2}  {\l}^{\frac \nu 2}   
\le[\K\, \frac {2^{-\nu}  u^\nu} {\Gamma(1+\nu) }   
 - \frac   
{2^{\nu} u^{-\nu} }{\Gamma(1-\nu)} \ri]\ .   
\label{asymmext}   
\ee    
As usual these functions do not belong to $L^2({\mathbb R} ^+, {\rm d} u)$ but any   
wave--packet does; moreover any such wave packet has this asymptotics.    
This allows us to find which parameter $\K$ corresponds to which      
unitary operator $e^{i\alpha}:H_+\mapsto H_-$, i.e. to a specific   
self--adjoint extension. Indeed, by matching the asymptotics in   
eqs. (\ref{asymext}) with that in Eq. (\ref{asymmext}) we obtain    
\bes   
\K = \frac{\cos\le(\frac \alpha 2-\frac {\pi\nu}4 \ri)}{\cos\le( \frac   
\alpha 2+ \frac {\pi\nu}4\ri)}\ .   
\ees    
We can now show  that the (``bulk-to-bulk'')     
two--point function (\ref{kgtp})   
in $AdS_{d+1}$ in the whole range $\nu\in (-1,\infty)$     
can be decomposed as follows:   
\bea   
&&    
W^{d+1}_{\nu}(Z( {v},z),Z'( {v'},z'))= \int_0^\infty d\l   
\theta_\l( {v})\theta_\l( {v}') W^{{{{M}}},d}_\lambda(z,z')\ , \    
\ \nu\in[1,\infty) \cr   
&&    
W^{d+1}_{\nu}(Z( {v},z),Z'( {v'},z')) =\int_0^\infty\!\!\!\!\! d\l   
\theta_\l^{(\infty)}( {v})\theta_\l^{(\infty)} ( {v}')   
W^{{{{M}}},d}_\lambda(z,z'), \    
 \nu\in [0,1) \cr   
&&    
W^{d+1}_{\nu}(Z( {v},z),Z'( {v'},z')) =\int_0^\infty d\l   
\theta_\l^{(0)}( {v})\theta_\l^{(0)}( {v}') W^{{{{M}}},d}_\lambda(z,z'), \    
 \nu\in (-1,0),\cr 
&&\    
\label{deco}   
\eea   
where $W^{{{{M}}},d}_\lambda(z,z')$ is the usual  two--point function   
for a Klein-Gordon field on $\mathbb M^d$ of square mass $\lambda$ in the   
Wightman vacuum:    
\bea   
&&W^{{{{M}}},d}_\lambda (z,z') \equiv \int \frac{{\rm d}^d p}{(2\pi)^{d-1}}   
\delta(p^2-\l)\Theta(p_0) e^{-ip\cdot(z-z')} = \cr   
&&\hspace {2cm}=    
(2\pi)^{-\frac d 2} \le(\frac {\delta }{\sqrt{\lambda}} \ri)^{\frac   
{2-d}2} K_{\frac {d-2}2}\le(\sqrt{\lambda} \delta  \ri)\ ;\qquad \delta\equiv
-(z-z')^2\ .   
\label{tpmink}
\eea   
In Eqs. (\ref{deco})  the functions $\theta_\l^{(\infty)}$  and the   
$\theta_\l^{(0)}$ belong to the domains of self--adjointness corresponding to   
the values  $\K=\infty$ and $\K=0$ respectively. They  explicitly read    
\bea   
&& \theta_\l^{(\infty)} ( {v}) = \frac 1{\sqrt {2}} e^{-\frac d 2  {v}   
} J_\nu(\sqrt{\l} e^{- {v}}) \label{modes>1}\\   
&& \theta_\l^{(0)} ( {v}) = \frac 1{\sqrt {2}} e^{-\frac d 2  {v}   
} J_{-|\nu|}(\sqrt{\l} e^{- {v}})\ .   
\eea   
The reason why we must use different self--adjoint extensions is that   
$W^{d+1}_\nu( Z({v},z) ,$ $ Z({v}',z') )$, as a function of   
$ {v}$ (or $ {v}'$) belongs to ${\mathfrak D}^{(\infty)}$ when   
$\nu\in[0,1)$ while it belongs  to ${\mathfrak D}^{(0)}$ when   
$\nu\in (-1,0)$:  
this can be  proved directly by studying the asymptotics.\par
The three Eqs. (\ref{deco}) are thus summarized into the following
formula valid for the whole range of parameter $\nu$:
\bea   
&& W^{d+1}_{\nu}(Z( {v},z),Z'( {v'},z'))=\cr
&& =(2\pi)^{-\frac d 2} ({u}\,{u}')^{\frac d 2}    
\int_0^\infty \frac{{\rm d}\lambda} 2 \, \lambda^{\frac {d-2} 4} J_\nu(\sqrt{\lambda} \,{u})   
J_\nu (\sqrt{\lambda}\,{u}')  K_{\frac {d-2}2} (m\,\delta)\ , 
\label{MAdS}   
\eea 
with, again,  $u=e^{-v}$.
The full  details of the proof    
 include analytical continuation to the Euclidean section where
$\delta=-(z-z')^2>0$, and take into
account formula (12) pag. 64 in \cite{Bateman2}.\\   
Eq. (\ref{deco}) can also been inverted and we obtain   
the Minkowski Klein-Gordon two-point function on the slice  $\Pi_v$   
by integrating $W_\nu$ against the eigenfunctions $\theta_\lambda$.   
For instance, when $\nu>1$ this corresponds to the introduction of the fields   
$\phi_{\lambda}(x)$ on the Minkowskian slice   
$\Pi_v$ obtained by smearing the AdS Klein-Gordon    
field $\PHI$  with  the complete   
set of    modes  (\ref{modes>1}):    
\be    
\phi_{\lambda } (x)= \int_{-\infty}^{\infty}    
\Phi(X(v,x)) \overline\theta_{\lambda} (v) e^{(d-2)v} dv .    
\label{hu}   
\ee     
It can be shown  that the field $\phi_{\lambda } (x)$    
is a canonical Minkowskian    
Klein--Gordon field in the Wightman vacuum   
state.    
In precise terms, we have that the AdS vacuum expectation value    
of $\phi_{\lambda } (x)$ is given by    
\be   
W_{\lambda,\lambda'}(x,x')\equiv    
\langle \Omega|\phi_\l(x)\phi_{\l'}(x')|\Omega\rangle = \delta(\l-\l')    
W^{{M},d}_\lambda(z,z').    
\label{restrizione}   
\ee   
 In particular,  the fields $\phi_\l$ have zero correlation    
(and hence commute) for different values of the square mass $\l$.\par\vskip 4pt 
The results of this section can be used to  construct other
two--point functions $W^{d+1,(\K)}_\nu$ $(Z(v,z),Z(v',z'))$
 for a  Klein--Gordon field  on AdS by using the other self--adjoint
extensions: however it is not guaranteed that such  $W^{d+1,(\K)}_\nu$ can be
extended to the other half of AdS since the definition uses the set of
coordinates defined only on one half. Moreover one should prove (or disprove)
the AdS invariance and analyticity properties of such states.
We will not go any further in this direction in this paper.  
  \noindent
\section{General QFT's in the Poincar\'e coordinates}

The results of section 4 and section 5 suggest the following alternative
approach to the AdS/CFT correspondence.
Starting from a given  set of AdS invariant $n$-point
 functions satisfying general
requirements of the form described in section 2,  
it is (at least formally) possible
 to obtain a set of Poincar\'e invariant (see below) $n$-point functions 
in one-dimension less by taking the following  restrictions:
\begin{equation}
{\mathcal W}_{n\{ v\}}^ M (x_1,\ldots, x_n)=
{\mathcal W}_n\left(X_1( {v},x_1), \ldots X_n( {v},x_n)\right).
\label{poinv}
\end{equation}
On the basis of the  requirement of asymptotic dimensionality c) 
supplemented by an argument 
similar to the one given
in section 4.2 (based on Eqs. (\ref{19}) and (\ref{Mink})) for justifying the
limit (\ref{55}) of two-point functions in the slices $\Pi_v$, 
$n-$point correlation functions on the boundary will be obtained
by taking the following limits:
\begin{equation}
{{\mathcal W}_n^{{{M}}}}
 (x_1,\ldots, x_n) = \lim_{ v \to \infty}
e^{nv\Delta} {\mathcal W}_{n\{ v\}}^ M (x_1,\ldots, x_n)
\label{adscft}
\end{equation}
One can also consider a many-leaf restriction as follows:
\begin{eqnarray}
&& {\mathcal W}_{n\{ {v}_{m+1},\ldots, {v}_n\}}
(X_1,\ldots, X_m, x_{m+1},\ldots, x_n)= \cr
&& = {\mathcal W}_n\left(X_1, \ldots X_m,X_{m+1}( {v}_{m+1},x_{m+1}),\ldots,X_{n}( {v}_n,x_n) \right),
\end{eqnarray}
and get various bulk-to boundary correlation functions by   
taking the limit as before:
\begin{eqnarray}
&&{\mathcal W_n}
 (X_1,\ldots, X_m, x_{m+1},\ldots, x_n) = \cr 
&& \lim_{ {v}_{m+1},\ldots, {v}_n \to \infty}
e^{(v_{m+1}+\cdots+v_n)\Delta} {\mathcal W}_{n\{ {v}_{m+1},\ldots, {v}_n\}}
(X_1,\ldots, X_m, x_{m+1},\ldots, x_n).\label{adscft1}
\end{eqnarray}
Restricting ourselves here to the limiting
procedure described by Eq. (\ref{adscft}), 
we then see that the general AdS/CFT correspondence for QFT's 
described in section 3 can alternatively be presented purely in terms  
of a limit of Minkowskian fields, denoted as follows:
\begin{equation}
\Phi(X) \to \{{\varphi}_v(x)\}  \to {\mathcal O}(x),
\label{Minklim} 
\end{equation}  
where each field ${\varphi}_v(x)$ 
is the scalar Minkowskian field whose $n-$point correlation functions 
are those given by (\ref{poinv}).

Here we must point out that there is a substantial difference
between two-point and  $n-$point functions.
In fact, in view of their maximal analyticity property  (see section 4.2) 
the two-point functions admit
restrictions to the slices $\Pi_v$ which are themselves boundary values of 
holomorphic functions in relevant  
Minkowskian
 complex domains of 
the corresponding complexified slices $\Pi_v^{(c)}$: in this  case 
there is therefore no problem of restriction 
of the distribution ${\cal W}_2$ to $\Pi_v \times \Pi_v$. 

As regards the $n-$point correlation functions, 
the existence of the restrictions  
(\ref{poinv}) as distributions on
 $(\Pi_v)^n$ is not an obvious consequence of the requirements
a), b), c) of section 2. 
Only the existence of the corresponding restrictions 
{ at Euclidean points} of 
$(\Pi_v^{(c)})^n$ 
(namely the Schwinger functions of these Minkowskian theories) are direct 
consequences of the spectral condition b) we have assumed: 
this is because  
changing $\tau$ into
 $i \sigma$ in (\ref{sphericcoordinates}) or changing $x^0$ into 
$i y^0$ in (\ref{coordinates}), all other parameters being kept real, yield two equivalent representations
of the Euclidean points of $\widehat {AdS}_{d+1}^{(c)}$.  

As a matter of fact, in order to be able to define the restrictions
(\ref{poinv}) as distributions 
enjoying the full  structure of Minkowskian $n-$point functions, namely as
distribution   
boundary values of holomorphic functions in relevant domains of 
$(\Pi_v^{(c)})^n$, one is led to use instead of b)
 an alternative  spectral condition 
 in which the positivity of the spectrum refers to 
the representation of a   $d-$dimensional
Abelian subgroup of $G$ playing the role of the Minkowskian
 translation group with respect to 
the slices $\Pi_v$. 

Let us briefly sketch   the construction.
Using the horocyclic parametrization of Eq. (\ref{coordinates}), 
we can lift the action of
the Poincar\'e\  group as follows.
Consider the standard action of the Poincar\'e group
on the Minkowski spacetime
 coordinates: $x'^\mu = \Lambda^\mu_\nu x^\nu
 + a^\mu$, $\mu=0,1,\ldots, d-1$. By
 plugging this relation into Eq. (\ref{coordinates})  we
promptly obtain the following relation:  
\be
\le\{\begin{tabular}{l}
$\displaystyle{X'^\mu = \Lambda^\mu_\nu X^\nu + (X^d+X^{d+1})a^\mu }$\cr
$\displaystyle{X'^d =\le(1+\frac {a^2}2 \ri) X^d +a_\mu \Lambda^\mu_\nu
X^\nu
+X^{d+1} \, \frac {a^2}2}$  \cr
$\displaystyle{X'^{d+1} 
 =\le(1-\frac {a^2}2 \ri) X^{d+1} -a_\mu\Lambda^\mu_\nu X^\nu  - 
X^{d} \,\frac {a^2}2}$ 
\end{tabular}
\right. ,
\label{ppp}
\ee
where Greek indices 
are raised and  lowered with the standard Minkowski metric. In matrix form we get 
\be
g(\Lambda, a) = \pmatrix{ 
 {{ {  \Lambda }}}  & a& a\cr
{\Lambda a }^{\, T} & \le(1+\frac {a^2}2\ri ) & \frac {a^2}2 \cr
-{\Lambda a} ^{\, T} & -\frac {a^2}2 &   \le(1-\frac {a^2}2\ri )
}
\ee
Among such transformations there is the Abelian subgroup of Poincar\'e\
translations $g(\mathbb I,a)$. 
The corresponding generators 
\be
P_\mu \equiv (X^d+X^{d+1}) \frac \pa{\pa X^\mu} + X_\mu\le( 
\frac \pa{\pa X^{d}}  - \frac \pa{\pa X^{d+1}}\ri)\ 
\ee
of these transformations form an Abelian algebra. 
The AdS spectral condition b) of section 2 should  then 
be supplemented by the following one: \\[10pt]
b$'$)  {\em Spectral condition: the infinitesimal generators $P^\mu$ 
are represented by (commuting)
 self-adjoint  operators whose joint spectrum is 
contained in the forward light-cone $V^{+} = \{p^\mu p_\mu \geq 0, p^{0} \geq 0\} $ of a 
$d$-dimensional Minkowski momentum space.\\[10pt]
By using a Laplace transform argument \cite{Streater,Reed},  
in the corresponding  vector variables $x_1,...,x_n$  
one can see that  this spectral condition implies 
the following analyticity property of the Wightman functions:
}\\[10pt]
{\em Analyticity corresponding to the spectrum of Poincar\'e translations: each  AdS 
distribution ${\mathcal W}_n\left(\! X_1(v_1,x_1)\! , \!  \ldots,\! X_{n}
( {v}_n,\! x_n)\!  \right)$
is the boundary value of a
holomorphic function ${W}_n (Z_1(v_1,z_1),$ $...,Z_n(v_n,z_n))$ 
which is defined in  the tube  
\begin{eqnarray}
{\mathcal T}_n = 
\{Z=(Z_1,...,Z_n) \in { {AdS}_{d+1}^{(c)}}; \ Z_j = Z_j(v_j,z_j);
\  v_1,\ldots,v_n \in \mathbb R,\ \cr
 \Im ({z_{j+1}-z_j})  \in V^+, \ j=1,\ldots,n-1 \}. 
\end{eqnarray} 
}\\[6pt] 
Property b$'$) implies in particular that it is meaningful
 to consider the restricted distributions  
${\cal W}^M_{n\{v\}}$ given in Eq. (\ref{poinv}).
The Poincar\'e invariance of  ${\cal W}^M_{n\{v\}}$ follows 
immediately by Eq. (\ref{ppp}).
 Furthermore,  the positive-definiteness of this family of distributions is induced as before by 
the analogous property satisfied   by  the distributions 
${\cal W}_n$ on 
$\widehat {AdS}_{d+1}$.
We also note that the validity of the Euclidean cluster property for
${W}^M_{n\{v\}}$ is equivalent to the condition introduced earlier in
Eq. (\ref{cluster}).
Under these conditions
the reconstruction procedure is now justified and  
the given field on  
${\widehat {AdS}_{d+1}}$ 
yields by restriction a well-defined quantum field $\varphi_v(x)$.

Moreover, it follows from the locality postulate a) 
together with Eqs. (\ref{lll})  
 that the field $\varphi_v$ also satisfies standard 
local commutativity  in $\Pi_v$.
Finally, in view of b$'$), the $n-$point functions of $\varphi_v$ 
are (for each $v$) boundary values of 
holomorphic functions  in the tube domains ${T}^M_n $ of Wightman's QFT. 
This  shows that these theories satisfy a 
standard energy-momentum spectrum 
condition (with respect to the generators
of spacetime translations). 
The conformal covariance of the boundary field $\mathcal O(x)$
 results from the general analysis of section 3.\\[10pt]

The interesting question whether the
 spectral condition b$'$) might be derived from condition b) 
together with AdS invariance 
will be left for future work.


\begin{thebibliography}{100}

\bibitem{Maldacena:1997re}
J. Maldacena,
\newblock Adv. Theor. Math. Phys. 2 (1998) 231, hep-th/9711200,
\newblock 

\bibitem{Gubser:1998bc}
S.S. Gubser, I.R. Klebanov and A.M. Polyakov,
\newblock Phys. Lett. B428 (1998) 105, hep-th/9802109,
\newblock 

\bibitem{Witten:1998qj}
E. Witten,
\newblock Adv. Theor. Math. Phys. 2 (1998) 253, hep-th/9802150,
\newblock 

\bibitem{Aharony:1999ti}
O. Aharony, S. S. Gubser, J. Maldacena, H. Ooguri and Y. Oz,
\newblock (1999), hep-th/9905111,
\newblock 

\bibitem{Balasubramanian:1998sn}
V. Balasubramanian, P. Kraus and A. Lawrence,
\newblock Phys. Rev. D59 (1999) 046003, hep-th/9805171,
\newblock 

\bibitem{Avis}
S.J. Avis, C.J. Isham and D. Storey,
\newblock Phys. Rev. D18 (1978) 3565,
\newblock 

\bibitem{Fronsdal:1974}
C. Fronsdal,
\newblock Phys. Rev. D10 (1974) 589,
\newblock 

\bibitem{Dirac}
P.A.M. Dirac,
\newblock Ann. Math. 36 (1935) 657,
\newblock 


\bibitem{rehren}
K.-H. Rehren,
\newblock (1999), hep-th/9905179.

\bibitem{haag}
R. Haag, 
\newblock Local Quantum Physics, (Springer--Verlag, Berlin, 1992),

\bibitem{Luscher:1975ez}
M. Luscher and G. Mack,
\newblock Commun. Math. Phys. 41 (1975) 203,
\newblock 

\bibitem{Streater}
R.F. Streater and A.S. Wightman,
\newblock PCT, spin and statistics, and all that (W. A. Benjamin, 1964).

\bibitem{Banks:1998dd}
T. Banks, M. R. Douglas, G. T. Horowitz and E. Martinec, 
\newblock (1998), hep-th/9808016,
\newblock 

\bibitem{Bros:1999}
J. Bros H. Epstein and U. Moschella,
\newblock In preparation .

\bibitem{Bros:1994dn}
J. Bros, U. Moschella and J. P. Gazeau, 
\newblock Phys. Rev. Lett. 73 (1994) 1746,
\newblock 

\bibitem{Bros:1996js}
J. Bros and U. Moschella,
\newblock Rev. Math. Phys. 8 (1996) 327, gr-qc/9511019,
\newblock 

\bibitem{Randall:1999ee}
L.~Randall and R.~Sundrum,
\newblock Phys.\ Rev.\ Lett.\  {\bf 83} (1999) 3370
[hep-ph/9905221].
  
\bibitem{Breitenlohner:1982jf}
P. Breitenlohner and D.Z. Freedman,
\newblock Ann. Phys. 144 (1982) 249,
\newblock 

\bibitem{Titchmarsh:1962}
E.C. Titchmarsh,
\newblock Eigenfunction expansion associated with second order differential
  equations (Clarendon Press, 1962).
\bibitem{Bros:1996bv2}
J.Bros, G. A. Viano,
\newblock Forum Math, 8 (1996) 659-722,

\bibitem{faraut}
J. Faraut, 
\newblock in Lect Notes in Math. 497, Springer-Verlag, Berlin 1975,

\bibitem{vilenkin}
N. Ja. Vilenkin, 
\newblock Fonctions Sp\'eciales et Th\'eorie de la Repr\'esentation des
groupes, (Dunod, Paris, 1969). 
 

\bibitem{O'Neill}
B. O'Neill,
\newblock Semi--Riemannian Geometry (Academic Press, 1983).

\bibitem{Hawking:1974df}
S.W. Hawking,
\newblock Commun. Math. Phys. 43 (1975) 199,
\newblock 

\bibitem{Gibbons:1977mu}
G.W. Gibbons and S.W. Hawking,
\newblock Phys. Rev. D15 (1977) 2738,
\newblock 

\bibitem{Sewell}
G.L. Sewell,
\newblock Ann. Phys. 141 (1982) 201.

\bibitem{Todorov}
G. Mack and I.T. Todorov,
\newblock Phys. Rev. D 8 (1973) 1764.

\bibitem{Bateman}
H. Bateman,
\newblock Higher Transcendental Functions (McGraw--Hill, 1954).

\bibitem{Klebanov:1999tb}
I.R. Klebanov and E. Witten,
\newblock (1999), hep-th/9905104,
\newblock 

\bibitem{Giddings:1999jq}
S.B. Giddings,
\newblock (1999), hep-th/9907129,
\newblock 

\bibitem{Bertola:1999ga}
M. Bertola, V. Gorini, U. Moschella and R. Schaeffer,
\newblock (1999), hep-th/9906035,
\newblock 

\bibitem{Reed}
M. Reed and B. Simon,
\newblock Methods of modern mathematical physics Vol. 2: Fourier analysis and
  self-adjointness (Academic Press, 1975).

\bibitem{Bateman2}
H. Bateman,
\newblock Tables of integral transforms (McGraw--Hill, 1954).

\end{thebibliography}
\end{document}